\documentclass{aa}
\usepackage{graphicx}
\usepackage{amsmath}
\usepackage{upgreek}
\usepackage{natbib}
\bibpunct{(}{)}{;}{a}{}{,}
\usepackage[usenames,dvipsnames]{color}
\usepackage{url}
\usepackage{hyperref}

\usepackage{xspace}
\usepackage{soul} 
\hypersetup{
    pdfborder={0 0 0 0},
    colorlinks=true,
    linkcolor=blue,
    urlcolor=blue,
    citecolor=blue
}

\def\fgl{1FGL~J1018.6$-$5856\xspace}

\begin{document}

\authorrunning{Marcote et al.}
\title{Refining the origins of the gamma-ray binary 1FGL~J1018.6$-$5856}
\titlerunning{Origin of 1FGL~J1018.6$-$5856}

\author{B.~Marcote\inst{\ref{inst1}}
\and M.~Rib\'{o}\inst{\ref{inst2},}\thanks{Serra H\'unter Fellow.}
\and J.~M.~Paredes\inst{\ref{inst2}}
\and M.~Y.~Mao\inst{\ref{instMao}}
\and P.~G.~Edwards\inst{\ref{instPhil}}
}

\institute{
Joint Institute for VLBI ERIC, Oude Hoogeveensedijk 4, 7991~PD Dwingeloo, The Netherlands. \email{marcote@jive.eu} \label{inst1}
\and
Departament de F\'isica Qu\`antica i Astrof\'isica, Institut de Ci\`encies del Cosmos (ICCUB), Universitat de Barcelona, IEEC-UB, Mart\'{\i} i Franqu\`es 1, E08028 Barcelona, Spain \label{inst2}
\and
Jodrell Bank Centre for Astrophysics, Alan Turing Building, School of Physics and Astronomy, The University of Manchester, Oxford Road, Manchester, M13 9PL, UK \label{instMao}
\and
CSIRO Astronomy and Space Science, Australia Telescope National Facility, PO Box 76, Epping, NSW 1710, Australia \label{instPhil}
}

\date{Received ... ; accepted ...}

\abstract{ 
Gamma-ray binaries are systems composed of a massive star and a compact object that exhibit emission from radio to very high energy gamma rays. They are ideal laboratories to study particle acceleration and a variety of physical processes that vary as a function of the orbital phase.
}{ 
    We aim to study the radio emission of the gamma-ray binary \fgl to constrain the emitting region and determine the peculiar motion of the system within the Galaxy to clarify its origin.
}{ 
    We analyzed an observation of \fgl with the Australian Long Baseline Array (LBA) at 8.4~GHz to obtain an accurate astrometry of the system and study its emission on milliarcsecond scales. We combined these data with the optical {\em Gaia} DR2 and UCAC4 catalogs to consolidate the astrometry information therein.
}{
    The gamma-ray binary \fgl shows compact radio emission ($< 3~\mathrm{mas}$ or $\lesssim 20~\mathrm{au}$ at $\sim 6.4~\mathrm{kpc}$ distance), implying a brightness temperature of $\gtrsim 5.6 \times 10^6~\mathrm{K}$, and confirming its nonthermal origin. We report consistent results between the proper motion reported by {\em Gaia} data release 2 (DR2) and the positions obtained from the {\em Gaia} DR2, UCAC4, and LBA data (spanning 20~yr in total). We also determined the distance to \fgl to be $6.4_{-0.7}^{+ 1.7}\ \mathrm{kpc}$.
Together with the radial velocity of the source we computed its three-dimensional (3D) proper and peculiar motion within the Galaxy. We obtained a peculiar motion of \fgl on its regional standard of rest (RSR) frame of $|u| = 45_{-9}^{+30}\ \mathrm{km\ s^{-1}}$, with the system moving away from the Galactic plane. In the simplest scenario of a symmetric stellar core collapse we estimate a mass loss of $ 4 \lesssim \Delta M \lesssim 9\ \mathrm{M_{\odot}}$ during the creation of the compact object.
}{
\fgl exhibits compact radio emission similar to that detected in other gamma-ray binaries. We provide the first accurate peculiar motion estimations of the system and place it within the Galaxy. The obtained motion and distance excludes the physical relation of the binary source with the supernova remnant (SNR)~G284.3$-$1.8.
}

\keywords{binaries : close - Gamma rays: stars - Radio continuum: stars - radiation mechanisms: nonthermal - stars: individual: 1FGL J1018.6$-$5856 - instrumentation: interferometers}

\maketitle

\section{Introduction}

The existence of emission at very high energies (VHE; $\gtrsim 100~\mathrm{GeV}$) requires the presence of powerful accelerators that speed particles up to relativistic energies. About two hundred sources have been found to exhibit VHE emission so far\footnote{See the online TeV Catalog \citep[TeVCat;][]{wakely2008}: \href{http://tevcat.uchicago.edu}{http://tevcat.uchicago.edu}.}. Most are associated with active galactic nuclei (AGNs), supernova renmants (SNRs), or pulsar wind nebulae (PWNe). However, a handful of these VHE sources are related to Galactic binary systems. With the exception of the colliding-wind binary $\upeta$-Carinae, recently detected up to 400~GeV \citep{leser2017}, all of these VHE binaries belong to the subclass of gamma-ray binary systems \citep[see e.g.,][and references therein]{dubus2013}. These binaries are ideal laboratories to study particle acceleration due to the involved short timescales of the emission, which is modulated by the orbital motion, and their relative proximity to us.

Gamma-ray binaries are composed of a massive star (typically of O or B spectral type) and a compact object, thought to be a young non-accreting neutron star in all cases. Gamma-ray binaries are distinguished from other high-energy binaries by the fact that they exhibit a spectral energy distribution (SED) dominated by the MeV--GeV photons, though emission is observed from radio to TeV $\upgamma$-rays. In these systems it is thought that a strong shock is produced between the stellar wind and the relativistic wind of the putative young non-accreting neutron star, generating a suitable environment to accelerate particles up to the required energies that subsequently produce the observed radiation \citep[see][for a detailed review]{dubus2013}.

Only seven gamma-ray binaries have been discovered to date: PSR~B1259$-$63 \citep{aharonian2005psr}, LS~5039 \citep{aharonian2005ls5039}, LS~I~+61~303 \citep{albert2006}, HESS~J0632+057 \citep{hinton2009,skilton2009}, 1FGL~J1018.6$-$5856 \citep{fermi2012}, LMC~P3 \citep[the first one discovered outside our Galaxy]{corbet2016}, and PSR~J2032+4127 \citep{lyne2015,veritasmagic2017}.
All of them are nonthermal radio emitters at gigahertz frequencies, with a spectrum dominated by synchrotron emission. All systems but one, LS~5039, show radio light curves modulated with the orbital motion \citep[see][]{dubus2013,marcote2015ls5039,marcote2016,marcote2015thesis}, and all gamma-ray binaries explored with very long baseline interferometric (VLBI) observations exhibit a compact, unresolved core plus an extended emission resolved on milliarcsecond scales. This extended emission exhibits morphological changes modulated by the orbital motion and typically represents $\sim$10\% of the total radio emission. It is usually interpreted as radiation emitted from the tail originated from the shocked material, although its origin remains unclear \citep[see e.g.,][]{moldon2012ls5039,moldon2012lspsr}.

\subsection{The gamma-ray binary 1FGL~J1018.6$-$5856}

The gamma-ray binary \textbf{}1FGL~J1018.6$-$5856\footnote{Also known as 3FGL~J1018.9$-$5856.}  was discovered as a periodic GeV source in the {\em Fermi}/LAT data \citep{abdo2010,abdo2011,corbet2011}. Variable nonthermal radio and X-ray counterparts, consistent with the position of a massive star, were later found by \citet{fermi2012}. Finally, a TeV source coincident with \fgl was reported by \citet{hess2012} also exhibiting periodic emission \citep{hess2015}.

The system is composed of an O6~V((f)) star and a compact object that orbits it every $16.544 \pm 0.008$~d \citep{napoli2011,fermi2012,an2015}. Assuming a total mass for the system of $\approx 29\ \mathrm{M_{\odot}}$ \citep{williams2015} we estimate a semi-major axis of the orbit of $\sim 0.4~\mathrm{au}$. The nature of the compact object remains unknown, although a $\sim 2$-$\mathrm{M_{\odot}}$ neutron star is favored \citep{strader2015, williams2015,waisberg2015,an2017,chen2017}.
The distance to \fgl has been determined to be $5.4_{-2.1}^{+4.6}~\mathrm{kpc}$ through photometric observations \citep{napoli2011}, while \citet{waisberg2015} suggest a distance $> 4~\mathrm{kpc}$ due to the observed extinction. We discuss the distance derived by the {\em Gaia} DR2 Catalog \citep{gaia2016,gaia2018} in Sect.~\ref{sec:distance}.

The radio emission of 1FGL~J1018.6$-$5856 is orbitally modulated with a quasi-sinusoidal variability that ranges between 1 and 6~mJy, and a flat spectrum of between 5 and 9~GHz \citep{fermi2012}. The maximum emission takes place at orbital phases of $\sim 0.3$, coincident with the maximum emission of the sinusoidal X-ray modulation. These light curves are however anticorrelated with the GeV and TeV ones \citep[see Fig.~3 in][]{hess2015}. More recent multifrequency radio observations covering the 2--33-GHz frequency range show that the orbital modulation decreases at higher frequencies, probably due to the presence of free-free absorption \citep[Coley et al.\ in prep]{coley2015}.

\subsection{The supernova remnant G284.3$-$1.8}

The gamma-ray binary \fgl is embedded in the Carina spiral arm, and is spatially coincident with the supernova remnant SNR G284.3$-$1.8 (also known as MSH~10$-$53).
This SNR was first discovered as a nonthermal radio source by \citet{mills1961} and \citet{milne1975}. Its emission is consistent with a $\sim 10^4$-$\mathrm{yr}$ old type-II SNR originated by a massive progenitor located at a distance of $\sim$2.9~kpc \citep{ruiz1986}, a value compatible at the $\sim 1$-$\sigma$ level with the aforementioned distance to \fgl. The expanding shock has interacted with molecular clouds located in the surrounding medium. \citet{milne1989} reported for the first time the full extent of SNR~G284.3$-$1.8 with observations made with the Molonglo Observatory Synthesis Telescope (MOST) and the Parkes Telescope, and concluded that all the detected radio emission has a nonthermal origin. TeV emission is detected in this region (in excess of \fgl's contribution), but all of it is probably produced by a PWN powered by the strong pulsar PSR~J1016$-$5857 (with a characteristic age of 21~kyr), located at the western edge of the mentioned SNR \citep{camilo2001,camilo2004}.
Given the spatial coincidence between SNR~G284.3$-$1.8 and \fgl, that their distances are compatible, and the age of the SNR, the possibility of a physical connection between both sources is not negligible. In that case, the binary system could likely be the progenitor of the SNR \citep[as suggested by e.g.][]{hess2012}.

In this paper we present the first VLBI observations of the gamma-ray binary \fgl, which have allowed us to put strong constraints on the compactness of its radio emission. The obtained astrometry, together with archival optical observations and the {\em Gaia} DR2 astrometry, provides the first accurate proper motion and distance measurements of the system. Furthermore, these data have allowed us to estimate the peculiar motion of the system within the Galaxy. We present the observations and the data reduction in Sect.~\ref{sec:obs}. We describe the obtained results and present the discussions in Sect.~\ref{sec:results}: on the observed radio emission of \fgl (Sect.~\ref{sec:disc1}), its proper motion (Sect.~\ref{sec:proper}) and its distance (Sect.~\ref{sec:distance}). Section~\ref{sec:disc2} describes the obtained results in the context of the Galaxy: its peculiar motion, its possible relation with SNR~G384.3$-$1.8 and the possible kick received in the system.
Finally, we present our conclusions in Sect.~\ref{sec:conclusions}.

\section{Radio observations and data reduction} \label{sec:obs}

\fgl was observed with the Australian Long Baseline Array (LBA) at 8.4~GHz on 26 April 2012 from 07:00 to 12:00~UTC (project code V454A; PI: Phil Edwards). The data were recorded with a total bandwidth of 32~MHz divided in four sub-bands of 16 channels each. A two-second integration time was used, recording full circular polarization. Seven stations participated in the observation: the phased Australian Telescope Compact Array (ATCA), Ceduna, Hobart, Mopra, Parkes, Warkworth, and one dish from the Australian Square Kilometre Array Pathfinder (ASKAP). Unfortunately, ASKAP could not produce reliable fringes and its data could not be used. Interferometric ATCA data were also recorded separately with a total bandwidth of 2~GHz (one sub-band of 33 channels) to be sensitive on much larger angular scales.

\begin{figure}
    \resizebox{\hsize}{!}{
        \includegraphics{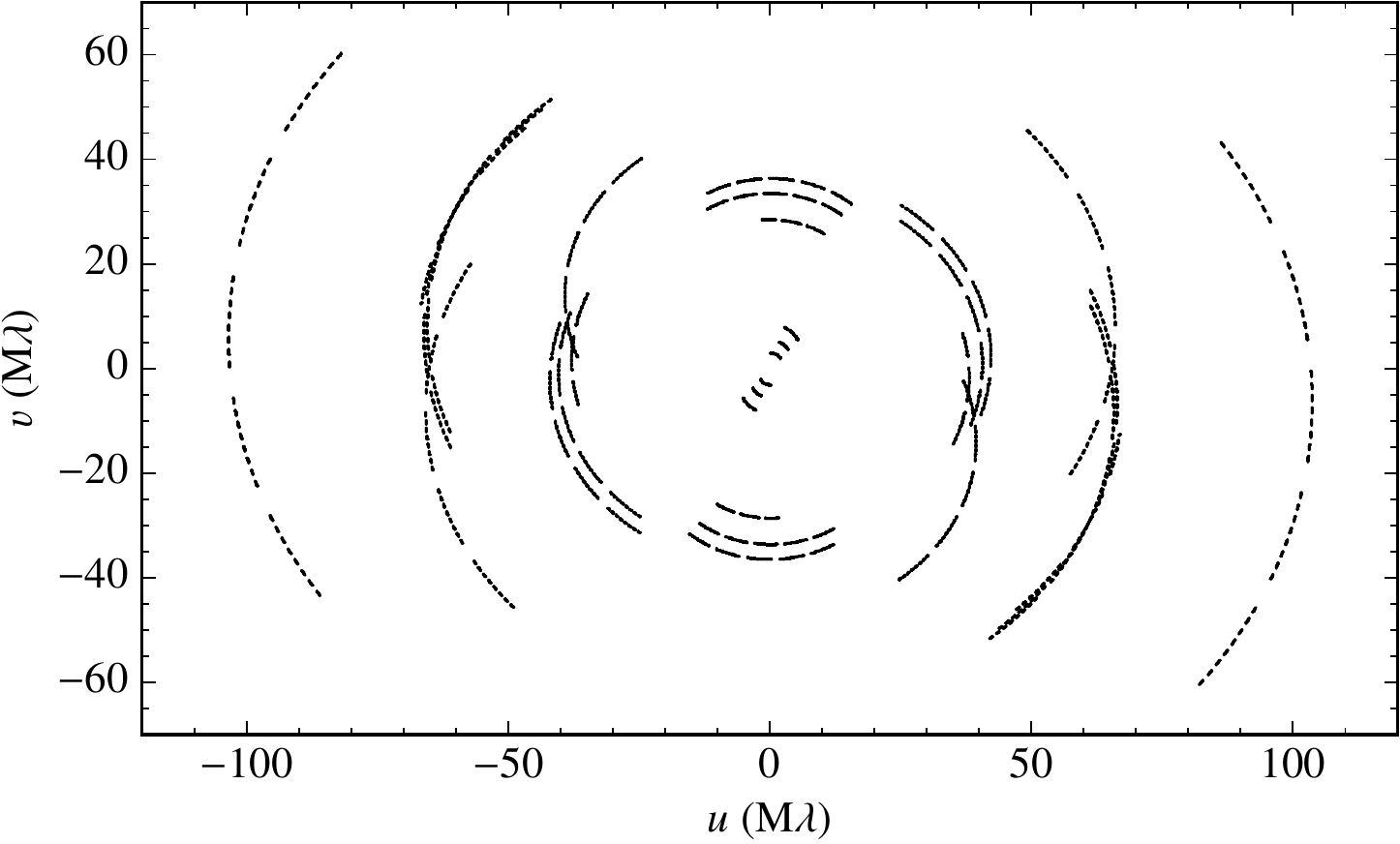}
    }
    \caption{Resulting $uv$-coverage for \fgl in the analyzed LBA observation. The central frequency is 8.4~GHz with a bandwidth of 32~MHz.}
    \label{fig:uvcoverage}
\end{figure}

The source 1057$-$797 was used as fringe finder and amplitude calibrator. J1019$-$6047 (located $\sim 1.9^{\circ}$ away from \fgl with a flux density of $\sim 0.4~\mathrm{Jy}$) was observed as phase calibrator in a phase-referencing cycle of 3~min on the calibrator and 3~min on the target source. 1FGL~J1018.6$-$5856 was thus observed during a total on-source time of 107~min. Figure~\ref{fig:uvcoverage} shows the obtained $uv$-coverage for 1FGL~J1018.6$-$5856 in this observation.

The LBA data have been reduced in AIPS\footnote{The Astronomical Image Processing System (AIPS) is a software package produced and maintained by the National Radio Astronomy Observatory (NRAO).} \citep{greisen2003} and Difmap \citep{shepherd1994} following standard procedures.
A-priori amplitude calibration was performed for each station using the system temperatures recorded during the observation, except for Warkworth where we used nominal system equivalent flux density (SEFD) values.
We flagged the bad visibilities manually and then fringe-fitted and bandpass calibrated the data using 1057$-$797 and J1019$-$6047.
We imaged and self-calibrated the phase calibrator, J1019$-$6047, to improve the solutions for each station. These solutions were then transferred to the target source, which was finally imaged. Due to the faintness of the emission we used a natural weighting during the imaging, which provides a higher sensitivity at the cost of a lower resolution.

The ATCA data were reduced in Miriad \citep{sault1995} and CASA\footnote{The Common Astronomy Software Applications, CASA, is also produced and maintained by the NRAO.} following standard procedures. We performed an amplitude calibration using 1057$-$797, which was observed by the ATCA Calibrator Database\footnote{\url{http://www.narrabri.atnf.csiro.au/calibrators/calibrator_database.html}} on 2012 April 22 and May 11 (four days before and 15 days after the LBA observation). After that we phase calibrated the data using J1019$-$6047. The obtained solutions were then transferred to \fgl, which was finally imaged using a natural weighting. We note that the lack of a calibrator with a well-known and constant flux density introduces an uncertainty in the absolute amplitude scale of the final images.

\section{Results and discussion} \label{sec:results}

\subsection{Radio emission of \fgl} \label{sec:disc1}

\begin{figure}
    \includegraphics[width=.51\textwidth]{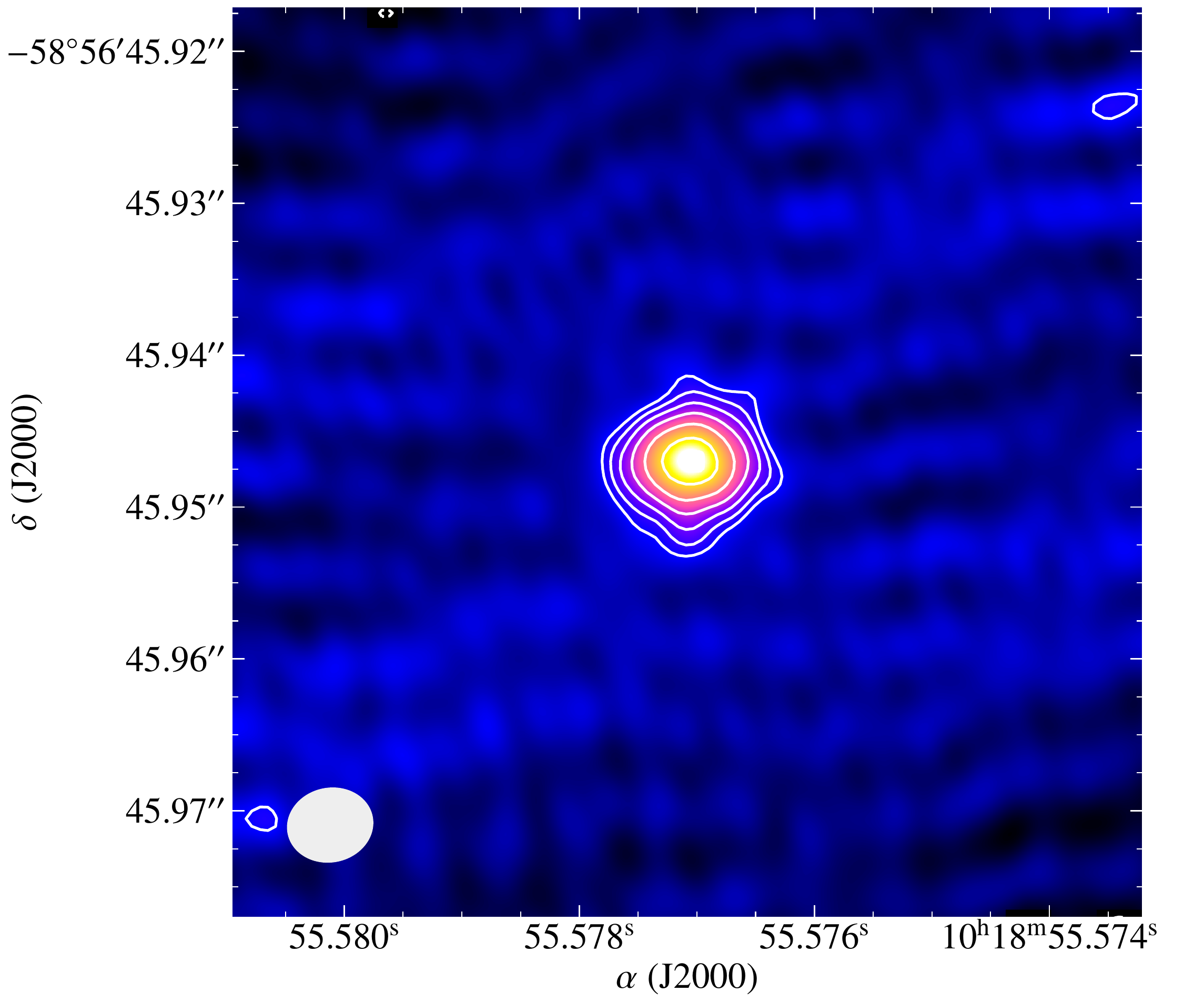}
    \caption{Image of \fgl obtained by the LBA at 8.5~GHz on 26 April 2012. Contours start at three times the rms noise level of $80~\mathrm{\upmu Jy\ beam^{-1}}$ and increase by factors of $\sqrt{2}$. The synthesized beam, shown by the gray ellipse at the bottom-left corner, is $4.8 \times 5.6~\mathrm{mas^2}$, $\mathrm{PA} = -78^{\circ}$.}
    \label{fig:image}
\end{figure}

\fgl is detected as a compact radio source on both arcsec and milliarcsecond scales from the ATCA and LBA data, respectively.
The ATCA data, with a synthesized beam of $44 \times 13~\mathrm{arcsec^2}$, $\mathrm{PA} = -6^{\circ}$, and a rms noise level of $90~\mathrm{\upmu Jy\ beam^{-1}}$, show a compact source with a flux density of $1.22 \pm 0.16~\mathrm{mJy}$.
The LBA data (see Fig.~\ref{fig:image}) show a compact emission of $2.01 \pm 0.15~\mathrm{mJy}$, where the rms noise level of the image is $80~\mathrm{\upmu Jy\ beam^{-1}}$ and the synthesized beam is $4.8 \times 5.6~\mathrm{mas^2}$, $\mathrm{PA} = -78^{\circ}$.

The obtained flux density values are slightly below but comparable to the values of $\sim 2$--$5~\mathrm{mJy}$ expected at that corresponding orbital phase ($\phi = 0.41$) from previous observations \citep{fermi2012,coley2015}. The fact that the ATCA data did not contain a calibrator with a well-known amplitude does not allow us to establish an accurate absolute amplitude calibration. We can thus only conclude that \fgl exhibits a flux density of $\sim 1$--$2\ \text{mJy}$. The presence of a large longer-term variability, such as in LS~I~+61~303, is in any case unlikely in \fgl.

By fitting a 2D Gaussian in the $uv$-plane we obtain an estimation of the source diameter of $\sim 3~\mathrm{mas}$ ($\sim 20~\mathrm{au}$ given the distance to the source, or around 50 times the semimajor axis of its orbit).
Whereas the measured size is only slightly smaller than the synthesized beam and is therefore significant \citep[see][]{martividal2012}, we cannot claim that the source is indeed resolved, due to the relatively weak phase calibrator ($\sim 0.4~\mathrm{Jy}$), its distance to \fgl (almost $2^{\circ}$ away), and the reduced number of stations producing a sparse $uv$-coverage (see Fig.~\ref{fig:uvcoverage}).
All these conditions can likely produce errors in the phase solutions transferred from the calibrator to the target source that affect the final calibration and produce an artificially larger source size.

\begin{table*}
        \caption{Position of \fgl as reported in the UCAC4 and {\em Gaia} DR2 catalogs, and the position measured from the LBA data. For each entry we provide the epoch of the observation, the right ascension ($\alpha$) and its uncertainty ($\Delta\alpha$), and the declination ($\delta$) and its uncertainty ($\Delta\delta$).}
    \label{tab:coordinates}
    \centering
        \begin{tabular}{l c l r@{}c@{}l l r@{}c@{}l}
                \hline\hline\\[-8pt]
            Catalog & Epoch & \multicolumn{1}{c}{$\alpha$} & \multicolumn{3}{c}{$\Delta\alpha$} & \multicolumn{1}{c}{$\delta$} & \multicolumn{3}{c}{$\Delta\delta$}\\
                & (yr) & & \multicolumn{3}{c}{(mas)} &  & \multicolumn{3}{c}{(mas)}\\[+2pt]
                \hline\\[-8pt]
        UCAC4 & 1997.74/1997.49\tablefootmark{a} & $10^{\rm h}18^{\rm m}55.588^{\rm s}$ & $13$&& & $-58^{\circ}56^{\prime}45.981^{\prime\prime}$ & $16$\\
                LBA & 2012.32 & $10^{\rm h}18^{\rm m}55.5771^{\rm s}$ & 1&.&4 & $-58^{\circ}56^{\prime}45.9474^{\prime\prime}$ & 0&.&8\\
                {\em Gaia} DR2 & 2015.5 & $10^{\rm h}18^{\rm m}55.5746^{\rm s}$ & 0&.&02 & $-58^{\circ}56^{\prime}45.94049^{\prime\prime}$ & 0&.&02\\
        \hline
    \end{tabular}
    \tablefoot{
        \tablefoottext{a}{The coordinates provided by UCAC4 refer to slightly different epochs for $\alpha$ and $\delta$, respectively.}
    }
\end{table*}
Comparing to the other explored gamma-ray binaries at VLBI scales \citep[see e.g.][]{moldon2012thesis}, we would expect extended emission on scales of $\sim 5~\mathrm{mas}$ with flux densities of $\sim 200~\mathrm{\upmu Jy\ beam^{-1}}$ (i.e., $\sim 10\%$ of the total flux density). This emission would not be significant above the rms noise level of our image. The obtained picture therefore shows that \fgl exhibits an emission which is consistent with respect to the other explored gamma-ray binaries.

The measured source size and flux density from the LBA image can additionally be used to set a lower limit on the brightness temperature of the source. In our case this size implies that $T_{\rm b} \gtrsim 5.6 \times 10^6~\mathrm{K}$, thus confirming the nonthermal origin for the radio emission.

The LBA data allowed us to measure the position of \fgl on milliarcsecond scales. We quote this position in Table~\ref{tab:coordinates} compared to the reported ones in the fourth U.S. Naval Observatory CCD Astrograph Catalog \citep[UCAC4,][]{zacharias2013} and the {\em Gaia} DR2 Catalog \citep{gaia2016,gaia2018}. The uncertainties in the quoted LBA position take into account the statistical uncertainties in the image (0.2~mas in both $\alpha$ and $\delta$), the systematic uncertainties associated with the uncertainty in the position of the phase calibrator \citep{beasley2002,gordon2016}, and the estimated errors due to the phase-referencing technique \citep{pradel2006}.

\subsection{Proper motion of \fgl} \label{sec:proper}

The position obtained for \fgl from the LBA observation, together with the positions reported by UCAC4 and {\em Gaia} DR2 (see Table~\ref{tab:coordinates}), allow us to measure the proper motion of the source. A weighted least-squares fit to the data provides a value of
\begin{eqnarray}
    \mu_{\alpha}\cos\delta \text{ (fit)} &=& - 5.9 \pm 0.4\ \mathrm{mas\ yr^{-1}}\nonumber\\
    \mu_{\delta} \text{ (fit)} &=& \phantom{-} 2.2 \pm 0.3\ \mathrm{mas\ yr^{-1}},
\end{eqnarray}
for a reference position at the epoch 2015.0 of
\begin{eqnarray}
    \alpha (\text{2015.0}) &=& \phantom{-}10^{\rm h}18^{\rm m}55.575066^{\rm s} \pm 0.17\ \text{mas}\nonumber\\
        \delta (\text{2015.0}) &=& -58^{\circ}56^{\prime}45.94180^{\prime\prime} \phantom{~~\,}\pm 0.16\ \text{mas.}
\end{eqnarray}

The obtained proper motion can be compared to the one reported by UCAC4 ($- 4.7 \pm 3.3,\ - 0.1 \pm 3.3\ \mathrm{mas\ yr^{-1}}$), which is one order of magnitude less precise, and to the one derived by {\em Gaia} DR2:
\begin{eqnarray}
    \mu_{\alpha}\cos\delta \text{ ({\em Gaia} DR2)} &=& - 6.41 \pm 0.05\ \mathrm{mas\ yr^{-1}}\nonumber\\
    \mu_{\delta} \text{ ({\em Gaia} DR2)} &=& \phantom{-} 2.21 \pm 0.05\ \mathrm{mas\ yr^{-1}},
\end{eqnarray}
for the reference position at the epoch 2015.5 quoted in Table~\ref{tab:coordinates}.

\begin{figure}
    \resizebox{\hsize}{!}{
        \includegraphics{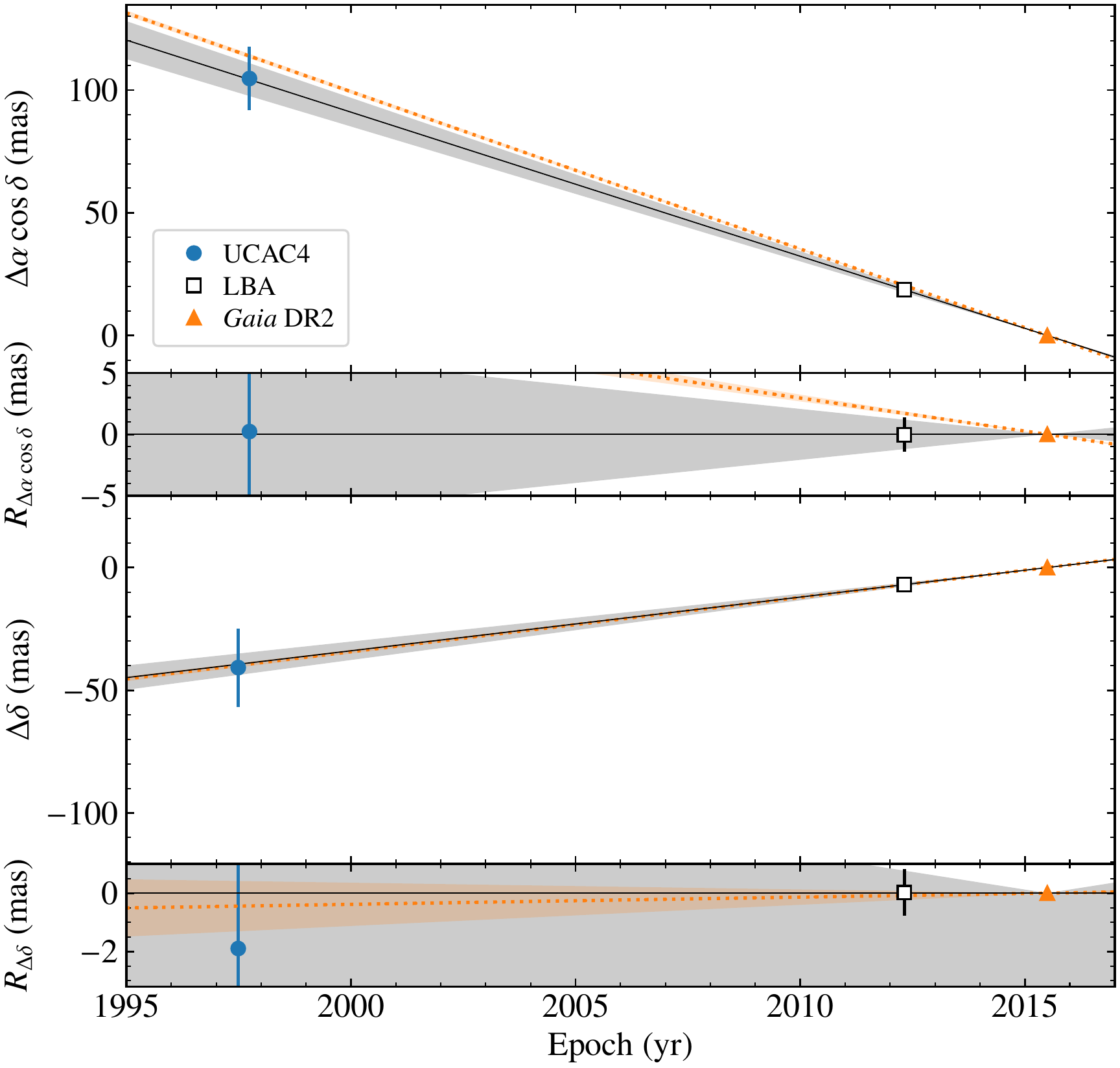}
    }
    \caption{Astrometry of \fgl from optical and radio data. The black line represents the proper motion fit to the data and the gray region its 1-$\sigma$ confidence interval. The dotted orange line represents the proper motion reported by {\em Gaia} DR2 and the orange region its 1-$\sigma$ confidence interval. $R$ represents the residuals from each fit. Error bars show the 1-$\sigma$ uncertainty on the individual positions. In the case of the {\em Gaia} DR2 data the error bars are smaller than the markers. Positions are referred to the {\em Gaia} DR2 position at J2015.5.}
    \label{fig:propermotion}
\end{figure}

Figure~\ref{fig:propermotion} shows the reported positions and proper motions during the covered period of time (almost 20~yr).
We note that both proper motions, and the {\em Gaia} DR2 prediction at the LBA epoch, agree within a $1.2$-$\sigma$ confidence level. From now on, we use the proper motion derived by {\em Gaia} DR2 due to its smaller uncertainty.

We note that, as shown in the following section, both the UCAC4 and LBA positions are not affected by the parallax and the orbital motion of \fgl, as the uncertainties in these positions are larger than these motions. We can therefore conclude that the inferred values from our fit are truly dominated by the proper motion of the system. Additionally, we note that the goodness of fit from the {\em Gaia} DR2 astrometry is $\sim 0.1$, confirming the reliability of the results.

\subsection{Distance to \fgl} \label{sec:distance}

\fgl appears in the {\em Gaia} DR2 Catalog with a parallax of $\varpi = 0.153 \pm 0.025\ \mathrm{mas}$.
The estimation of the distance to the source from a parallax measurement is not trivial \citep{bailerjones2015,luri2018}, and we have considered it as an inference problem where the parallax measurement is described by a Gaussian probability distribution function. We assume a prior $\mathcal{P}_\text{prior} (D) = D^2 \exp ( - D/L )$, where $D$ is the distance and $L$ is the length scale. This prior corresponds to an exponentially decreasing volume density, and is optimal in our case due to its simplistic but physical meaning (see the aforementioned references for a detailed discussion).
We have set $L$ to be the previous estimation on the distance to \fgl ($\sim 5.4\ \mathrm{kpc}$; \citealt{napoli2011}), as it represents the a-priori most probable distance from our prior.

\begin{figure}
    \resizebox{\hsize}{!}{
        \includegraphics{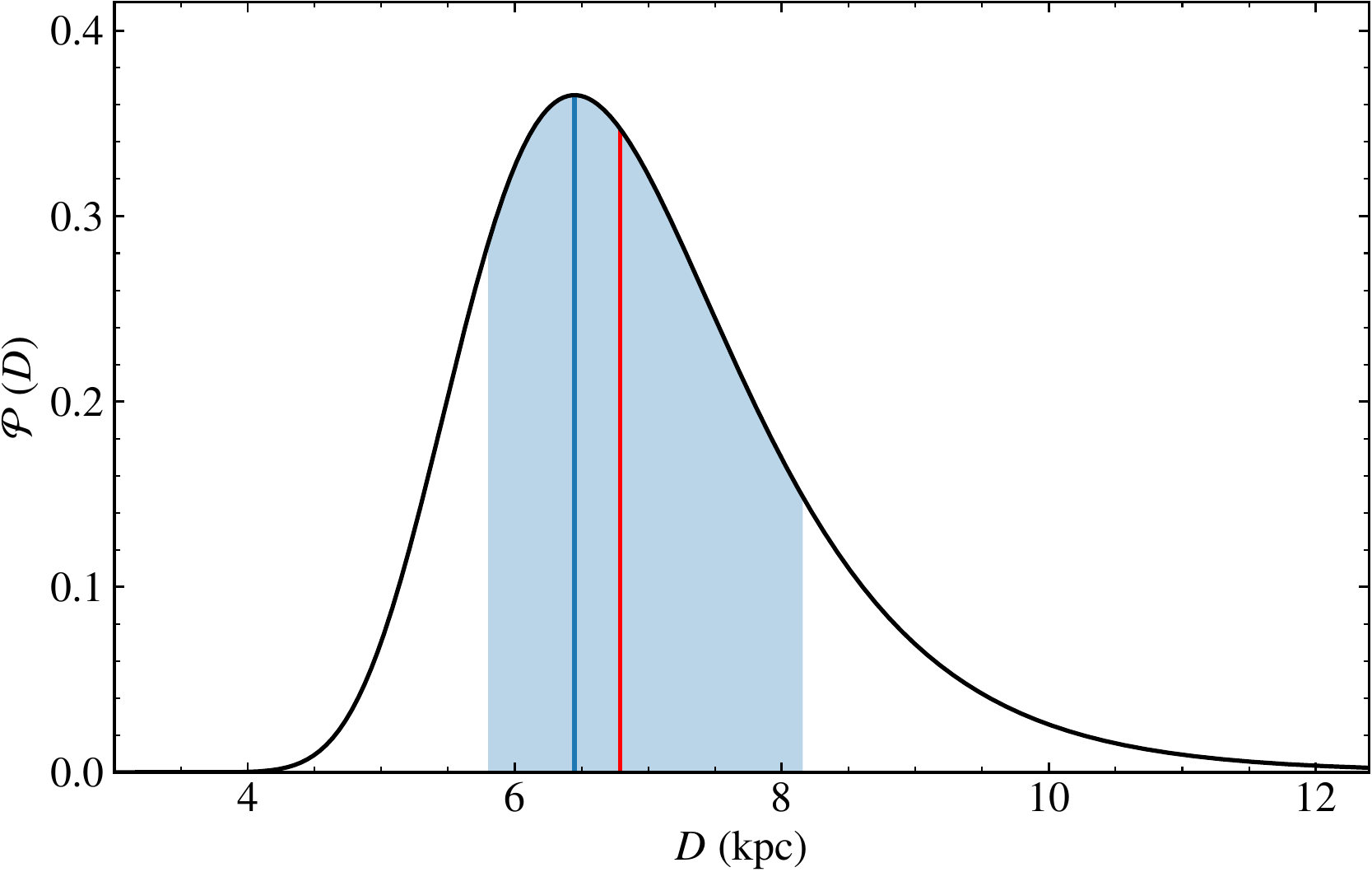}
    }
        \caption{Probability distribution function, $\mathcal{P}$, of the distance $D$ to \fgl derived from the {\em Gaia} DR2 parallax by assuming a prior $\mathcal{P}_\text{prior} (D) = D^2 \exp (- D/L)$ (see text for more details). The blue vertical line represents the mean distance to the source and the light blue shadowed region represents its $\pm 1\sigma$ confidence interval ($D = 6.4_{-0.7}^{+ 1.7}\ \mathrm{kpc}$). The red vertical line represents the median distance (6.8~kpc).}
    \label{fig:prob}
\end{figure}
Using the {\em Gaia} DR2 parallax we derive a mean distance to \fgl of $D = 6.4_{-0.7}^{+ 1.7}\ \mathrm{kpc}$, where the median distance is $6.8\ \mathrm{kpc}$ (see Fig.~\ref{fig:prob}). From now on we use the derived mean value as the distance to \fgl. We note that the use of different priors (varying $L$ and/or using constant probability functions) produce consistent results (within the reported uncertainties for all reasonable cases).

For consistency we have determined that, at the derived distance, the orbit of \fgl exhibits an angular size of $\approx 0.12\ \mathrm{mas}$ (for a semi-major axis of $\approx 0.4~\mathrm{au}$). This value is lower than our precision in the LBA results. In the case of the {\em Gaia} results, we note that the optical emission only arises from the companion star, and therefore these results are exclusively sensitive to the motion of this component, which only represents $\approx 9\ \mathrm{\upmu as}$ (for the masses reported in the introduction). This value is a factor of two lower than the uncertainties in the {\em Gaia} results. We can therefore confirm that the orbital motion of the system does not significantly affect the derived proper motion or parallax.

\subsection{\fgl within the Galaxy} \label{sec:disc2}

In the following we discuss the implications of all the obtained results on the motion of the source within the Galaxy (Sect.~\ref{sec:propermotion}) and with respect to SNR~G284.3$-$1.8 (Sect.~\ref{sec:relationsnr}). We also estimate the mass loss required to produce the observed peculiar motion in Sect.~\ref{sec:kick}.

\begin{figure}
    \resizebox{\hsize}{!}{
        \includegraphics{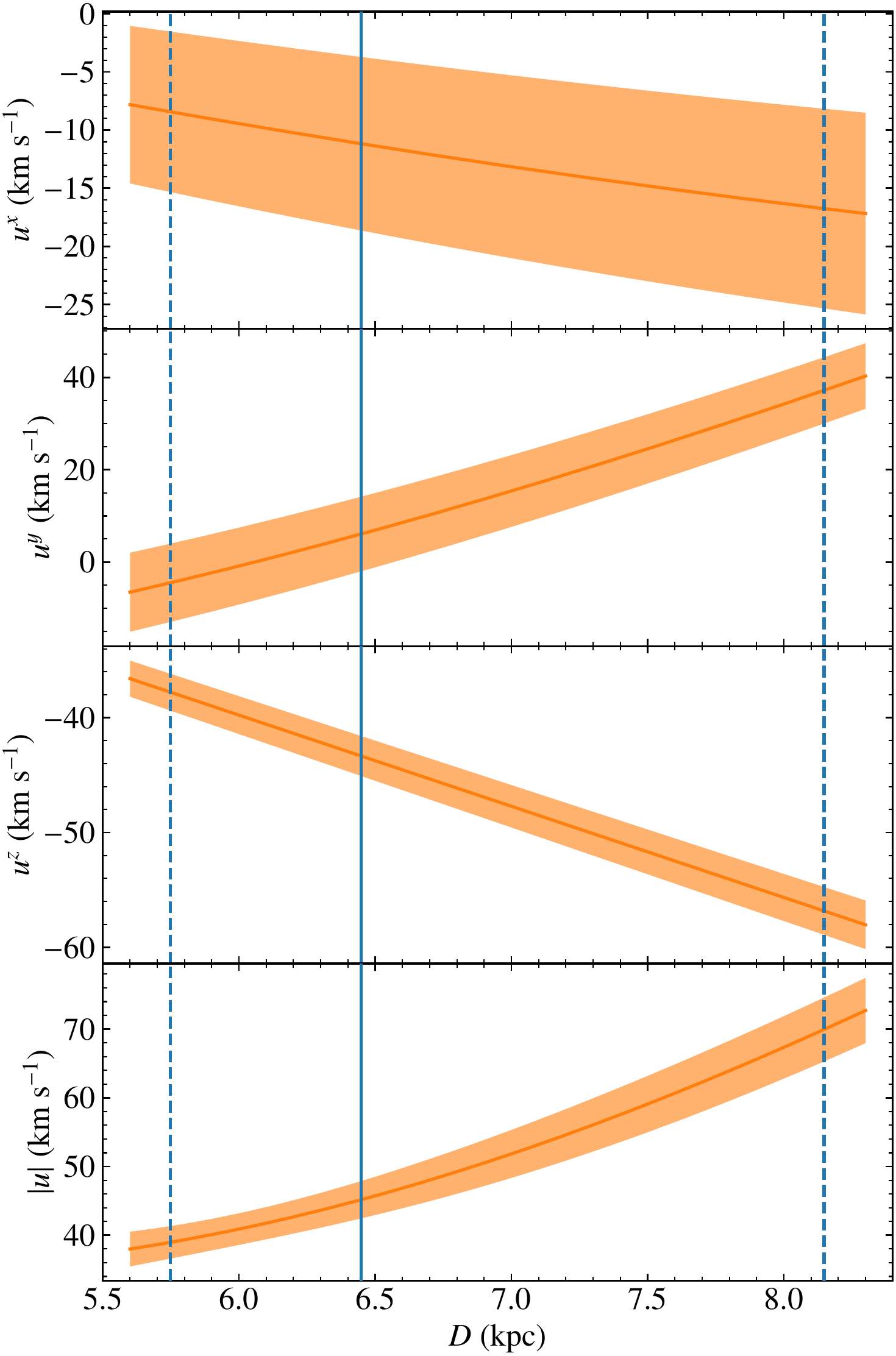}
    }
    \caption{Peculiar motion of \fgl on its regional standard of rest (RSR) as a function of the distance between the Sun and the source. The orange lines show the average velocity whereas the colored region represents the 1-$\sigma$ confidence interval. We show the components in the direction to the Galactic center ($u^x$), to the Galactic rotation ($u^y$), and towards the Galactic north pole ($u^z$). The bottom panel shows the module of the peculiar motion ($|u|$). The vertical blue lines represent the average distance to \fgl (solid line) and its $\pm 1$-$\sigma$ confidence interval (dashed lines).}
    \label{fig:velocities}
\end{figure}

\subsubsection{Peculiar motion} \label{sec:propermotion}

Transforming the {\em Gaia} DR2 proper motion to Galactic coordinates we obtain a Galactic proper motion of $\mu_{l} = -6.57 \pm 0.05~\mathrm{mas\ yr^{-1}} \text{ and } \mu_{b} = -1.68 \pm 0.05~\mathrm{mas\ yr^{-1}}$. Given that \fgl is located at a negative Galactic latitude ($-1.69^{\circ}$), this proper motion implies that the source is going away from the Galactic plane. Given the estimated distance to the source of $D = 6.4_{-0.7}^{+1.7}~\mathrm{kpc}$ the measured proper motion implies a tangential velocity of $v_{\rm t} \sim 210_{-20}^{+50}~\mathrm{km\ s^{-1}}$.

\begin{figure}
    \resizebox{\hsize}{!}{
        \includegraphics{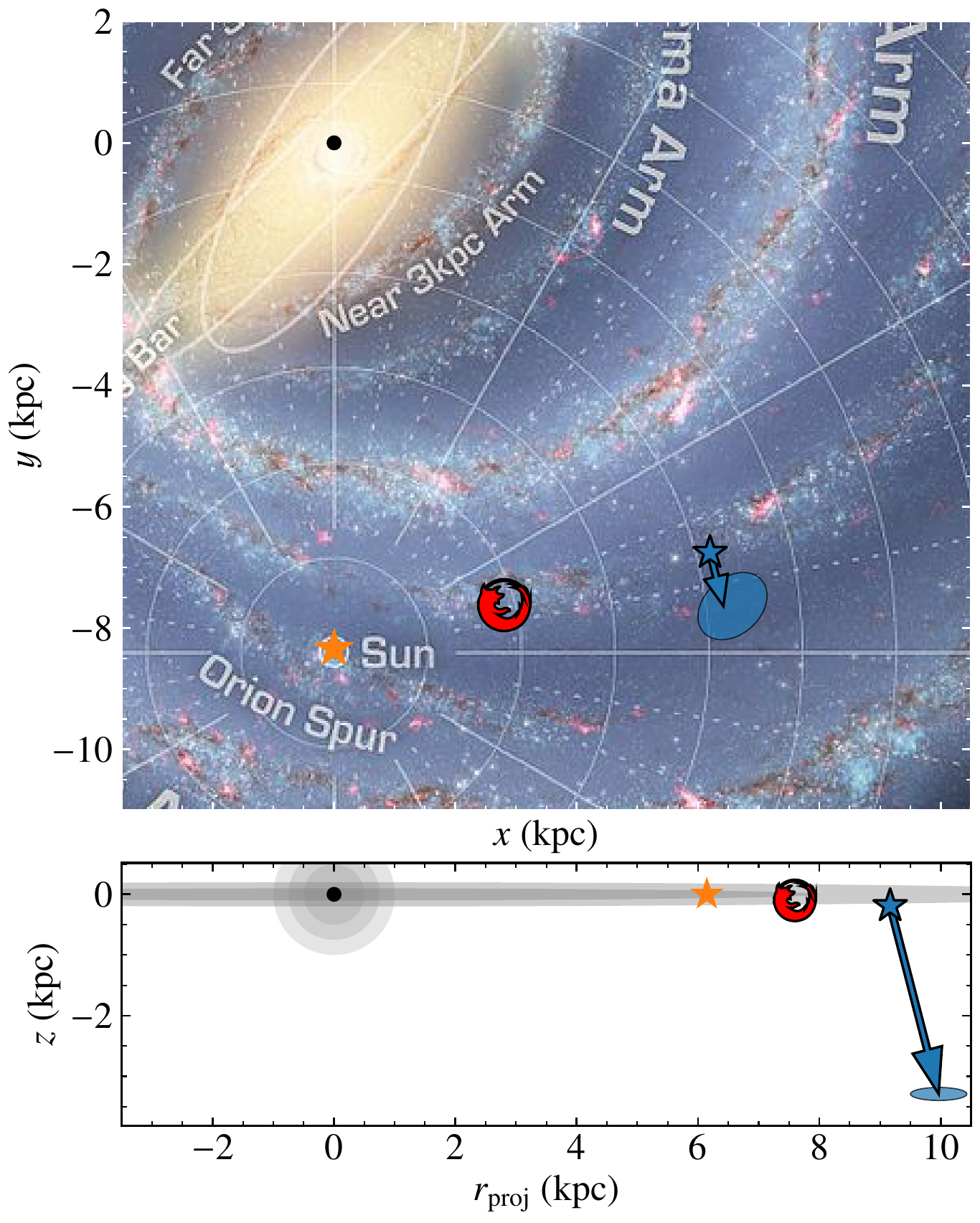}
    }
    \caption[\color{red}Caption for LOF]{Absolute position and peculiar motion of \fgl within the Galaxy in a view perpendicular to the Galactic plane ({\em top}) and parallel to it in the projected direction between the Galactic center and \fgl ({\em bottom}).  The black dot represents the Galactic center, the orange star represents the Sun, the open red symbol represents SNR~G384.3$-$1.8, and the blue star represents \fgl at the distance of 6.4~kpc. The blue arrow represents the peculiar motion of the source (in the RSR, after Galactic rotation subtraction) after 70~Myr for reference\protect\footnotemark. The blue ellipse represents the uncertainty on this motion. The gray ellipses provide a schematic picture of the Galactic plane. {\em Image Credit:} NASA/JPL-Caltech/R.\ Hurt/SSC/Caltech, The Mozilla Foundation.}
    \label{fig:milky-way}
\end{figure}
We can combine these measurements together with the average radial velocity of the source of $v_{\rm r} \approx 30$--$36~\mathrm{km\ s^{-1}}$ \citep{strader2015} to determine the peculiar motion of \fgl within the Galaxy, on its regional standard of rest \citep[RSR;][]{reid2014}. We have assumed a distance to the Galactic center of $8.34 \pm 0.16$~kpc, the local standard of rest (LSR) defined by $V_{\odot} = (V_{\odot}^{X}, V_{\odot}^{Y}, V_{\odot}^{Z}) = (10.7 \pm 1.8,\ 15.6 \pm 6.8,\ 8.9 \pm 0.9)\ \mathrm{km\ s^{-1}}$, and a Galactic rotation of $\Theta_0 = 240 \pm 8\ \mathrm{km\ s^{-1}}$ \citep[see model A5 in][]{reid2014}, where $X,Y,Z$ are defined as $X$ towards the Galactic center, $Y$ towards $l = 90^{\circ}$ (following the Galactic rotation), and $Z$ towards the north Galactic pole.
Using this frame we obtain a peculiar motion in the RSR of $u = (u^x, u^y, u^z) = (-11_{-14}^{+10},\ 6_{-20}^{+40},\ -43_{-17}^{+8})\ \mathrm{km\ s^{-1}}$, where $x,y,z$ are defined as the RSR frame from \fgl to the Galactic center, to the Galactic rotating direction, and towards the Galactic north pole, respectively. Figure~\ref{fig:velocities} shows these results as a function of the distance $D$ to \fgl, and Fig.~\ref{fig:milky-way} shows the obtained peculiar motion of the source with respect to the Sun and the Galaxy.

\footnotetext{We note that the expected lifetime of an O6~V star is $\lesssim 5\ {\rm Myr}$ \citep{massey2001} and therefore at the quoted time such a  star will no longer exist.}
\begin{figure}
    \includegraphics[width=0.51\textwidth]{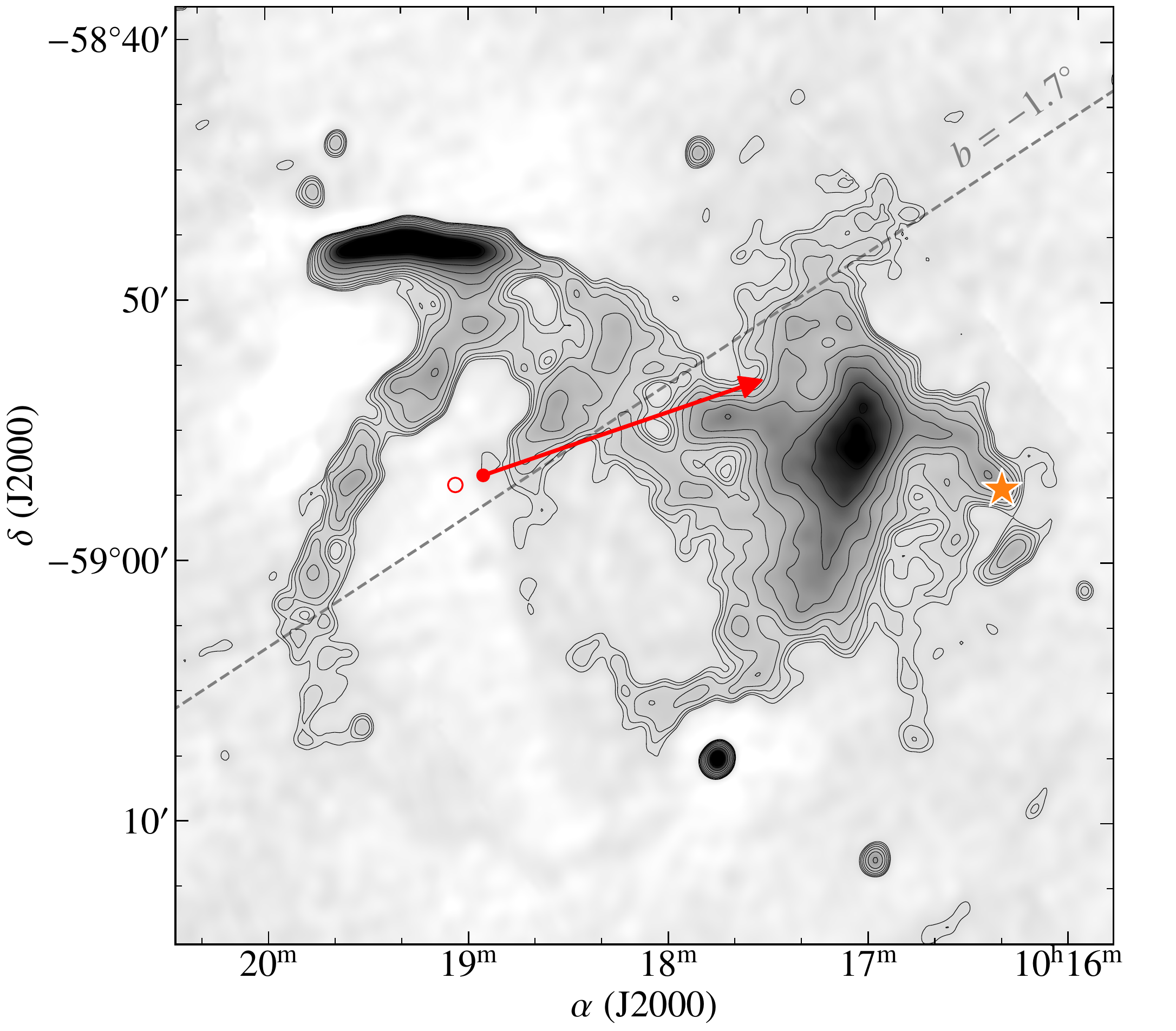}
    \caption{Field of \fgl as seen by the Sydney University Molonglo Sky Survey (SUMSS) at 843~MHz. Contours start at two times the rms noise level of 2~mJy with increments of $\sqrt{2}$. The emission in the region is dominated by the SNR~G284.3$-$1.8 (delimited by the east and west arcs) and its interaction with molecular clouds (e.g. the north-east strong filament). The filled red dot represents the current position of \fgl, while the empty red dot represents the position of the source $10~\mathrm{kyr}$ ago (when the supernova that created SNR~G284.3$-$1.8 took place). The red arrow represents the proper motion of the binary system and the predicted position in $100~\mathrm{kyr}$. The orange star in the west denotes the position of the pulsar PSR~J1016$-$5857 and its associated PWN.}
    \label{fig:snr}
\end{figure}

In summary, we obtain an absolute peculiar motion of $36 < |u| < 75\ \mathrm{km\ s^{-1}}$ with an average value of $\approx 45~\mathrm{km\ s^{-1}}$. We note that the final uncertainties quoted on the peculiar motion are dominated by the uncertainty on the distance of the source.
OB stars exhibit typical peculiar motions of $\lesssim 10~\mathrm{km\ s^{-1}}$ \citep[see e.g.,][]{melnik2017}, although these values can increase up to $10$--$30~\mathrm{km\ s^{-1}}$ in some regions \citep[see e.g.,][for the OrionBN/KL region]{gomez2008}. Therefore it is unclear if the derived peculiar motion originates from a kick during the stellar core collapse or if it was intrinsic to the system beforehand.
We note that the derived motion is similar to the motion of $26 \pm 8~\mathrm{km\ s^{-1}}$ observed in the gamma-ray binary PSR~B1259$-$63 \citep{millerjones2018}, but lower than the motion of $\sim 140~\mathrm{km\ s^{-1}}$ observed in LS~5039 \citep{moldon2012lspsr}, and higher than the motion of $\sim 16~\mathrm{km\ s^{-1}}$ observed in LS~I~+61~303 \citep{wu2018}. In any case, these motions are well below the velocities observed in isolated pulsars \citep[see e.g.,][]{chatterjee2005}, as expected for binary systems.

\subsubsection{Relation with SNR~G384.3$-$1.8} \label{sec:relationsnr}

\fgl is located near the center of the $10^4$-$\mathrm{yr}$ old SNR~G284.3$-$1.8. Figure~\ref{fig:snr} shows the field around \fgl as seen by the Sydney University Molonglo Sky Survey (SUMSS; \citealt{mauch2003}) at 843~MHz. We also show the position of \fgl at the time when the SNR was produced ($\sim 1~\mathrm{arcmin}$ to the SE from its current position) and its direction of motion. Figure~\ref{fig:milky-way} shows the 3D position of both the SNR and \fgl with respect to the Sun.

Although the distance to SNR~G384.3$-$1.8 ($\sim 2.9\ \mathrm{kpc}$) was initially compatible at $\sim 1$-$\sigma$ confidence level with the distance to \fgl reported in the literature ($5.4_{-2.1}^{+4.6}\ \mathrm{kpc}$), the distance derived in this paper ($6.4_{-0.7}^{+1.7}\ \mathrm{kpc}$) rejects the possible association of SNR~G384.3$-$1.8 with the core collapse that led to the compact object in \fgl at a 5-$\sigma$ confidence level. We note that \citet{ruiz1986} constrained the distance to SNR~G384.3$-$1.8 to be $\lesssim 3.5\ \mathrm{kpc}$ (with the aforementioned mean value of 2.9~kpc) due to the strong absorption expected in that line of sight.

Additionally, the obtained proper motion for \fgl is not compatible with a source moving away from SNR~G284.3$-$1.8 (see Fig.~\ref{fig:milky-way}), and in any case the binary system could not have traveled the distance of $\gtrsim 1\ \mathrm{kpc}$ in only $10^4\ \mathrm{yr}$ (this would imply velocities of $\gtrsim 10^5\ \mathrm{km\ s^{-1}}$, which would be astonishing for such a system).

We therefore conclude that \fgl and SNR~G284.3$-$1.8 are completely unrelated systems and do not share any common origin or physical connection. We note that in the vicinity of \fgl we see no other sources that could be linked to the binary system.

\subsubsection{Kick after the stellar core collapse} \label{sec:kick}

As mentioned, \fgl exhibits a peculiar motion slightly (but not significantly) larger than typical OB stars . It is unclear, but likely, that the stellar core collapse that formed the compact object produced a kick in the system and changed its peculiar motion to some extent.
We can only provide rough estimations of how much mass was lost from the system during the collapse under the assumptions of a completely symmetric stellar core collapse and that the observed peculiar motion is dominated by the received kick. Under these assumptions the velocity of the center of mass of the binary system would have changed after the collapse by a value of \citep{nelemans1999}
{\small
\begin{equation}
    \frac{\Delta M}{\rm M_{\odot}} = \frac{1}{213} \left(\frac{v_s}{\rm km\, s^{-1}}\right) \left(\frac{M_{\ast}}{\rm M_{\odot}}\right)^{-1} \left( \frac{P_{\text{re-circ}}}{\rm d} \right)^{1/3} \left( \frac{M_{\rm X} + M_{\ast}}{\rm M_{\odot}} \right)^{5/3}
,\end{equation}}
\hspace{-3pt}where $\Delta M$ is the lost mass in the collapse (and therefore ejected from the binary system), $v_s$ is the velocity of the binary system due to the kick, $M_{\ast}$ is the initial mass of the companion star, $P_{\rm re-circ}$ is the orbital period after re-circularization after the collapse, and $M_{\rm X}$ is the remaining mass of the compact object.
$M_{\rm X}$ is estimated to be $\sim 2~\mathrm{M_{\odot}}$ (as discussed in the introduction). Given that the total mass of the system is $\approx 29~\mathrm{M_{\odot}}$, the mass of the companion star must be $M_{\ast} \sim 27~\mathrm{M_{\odot}}$. $P_{\text{re-circ}} = P_{\rm orb} (1 - e^2)^{3/2}$ \citep{nelemans1999}, where $P_{\rm orb}$ is the final orbital period of the system (after the core collapse), and $e$ the final eccentricity of its orbit. Given that $e \sim 0.31 \pm 0.16$ \citep{monageng2017} we obtain $P_{\text{re-circ}} \approx 14.2\ \mathrm{d}$, and therefore $4 \lesssim \Delta M \lesssim 9\ \mathrm{M_{\odot}}$ for the derived peculiar motion of the source.

These values imply that a non-negligible mass was ejected from the system during the core collapse. Considering this lost mass, the progenitor would have a mass of $\sim 6$--$11\ \mathrm{M_{\odot}}$, which is smaller than the $\sim 27$-$\mathrm{M_{\odot}}$ companion star. These results suggest that the progenitor lost additional mass before the core collapse, either by mass transfer to the companion star or pulling away from the system via the stellar wind.
We note that the lost mass could easily be higher (likely $\sim 2$ times larger given the typical peculiar motions of such systems) if the system has changed the direction of motion during the core collapse.

\section{Conclusions} \label{sec:conclusions}

We report the first VLBI observations of \fgl, showing that the source exhibits compact emission on milliarcsecond scales. Its compactness implies a brightness temperature of $\gtrsim 5.6 \times 10^6\ \mathrm{K}$, confirming the nonthermal origin of the emission. Extended emission on these angular scales is not detected but the current constraints are consistent with a faint emission similar to the one reported in other gamma-ray binaries.
An ongoing monitoring of \fgl with more sensitive and with LBA observations of a  slightly higher
resolution would allow us to detect the presence of extended emission originating from the putative cometary tail as observed in other gamma-ray binaries, and study its morphological evolution along the orbit.

The presented LBA observation has allowed us to obtain a precise position of the source that, together with archival optical UCAC4 observations and the {\em Gaia} DR2 astrometry, provides the most accurate measurements of the proper and peculiar motions and distance to \fgl so far.
The derived results agree with the proper motions reported by {\em Gaia} DR2, providing an independent verification for this system. Given the distance and the characteristics of \fgl, we confirm that the orbital motion does not significantly affect the astrometry results. We also conclude that \fgl cannot physically be related to SNR~G284.3$-$1.8.

The obtained peculiar motion of \fgl is slightly, but not significantly, faster than the typical motions observed for early-type stars. The system thus did not receive a significant kick after the core collapse that led to the formation of compact object, although a significant amount of mass of $4\text{--}9\ \mathrm{M_{\odot}}$ was lost during this collapse.

\begin{acknowledgements}

The Australia Telescope Compact Array and the Australian Long Baseline Array are part of the Australia Telescope National Facility which is funded by the Australian Government for operation as a National Facility managed by CSIRO. This paper includes archived data obtained through the Australia Telescope Online Archive (\url{http://atoa.atnf.csiro.au}). This work has made use of data from the European Space Agency (ESA) mission {\it Gaia} (\url{https://www.cosmos.esa.int/gaia}), processed by the {\it Gaia} Data Processing and Analysis Consortium (DPAC, \url{https://www.cosmos.esa.int/web/gaia/dpac/consortium}). Funding for the DPAC has been provided by national institutions, in particular the institutions participating in the {\it Gaia} Multilateral Agreement. BM, MR, and JMP acknowledge support from the Spanish Ministerio de Econom\'ia y Competitividad (MINECO) under grants AYA2016-76012-C3-1-P, FPA2017-82729-C6-2-R, and MDM-2014-0369 of ICCUB (Unidad de Excelencia ``Mar\'ia de Maeztu'').

This research made use of APLpy, an open-source plotting package for Python hosted at \url{http://aplpy.github.com}, Astropy, a community-developed core Python package for Astronomy \citep{astropy2013}, and Matplotlib \citep{hunter2007}.

\end{acknowledgements}

\bibliographystyle{aa}
\bibliography{/jop93_0/Documents/Reference/bibliography.bib}

\newcommand{\SortNoop}[1]{}
\begin{thebibliography}{61}
\expandafter\ifx\csname natexlab\endcsname\relax\def\natexlab#1{#1}\fi

\bibitem[{{Abdo} {et~al.}(2010){Abdo}, {Ackermann}, {Ajello}, {Allafort},
  {Baldini}, {Ballet}, {Barbiellini}, {Bastieri}, {Bechtol}, {Bellazzini},
  {Berenji}, {Blandford}, {Bonamente}, {Borgland}, {Bouvier}, {Brandt},
  {Bregeon}, {Brez}, {Brigida}, {Bruel}, {Buehler}, {Burnett}, {Caliandro},
  {Cameron}, {Caraveo}, {Carrigan}, {Casandjian}, {Cecchi}, {{\c C}elik},
  {Chaty}, {Chekhtman}, {Cheung}, {Chiang}, {Ciprini}, {Claus}, {Cohen-Tanugi},
  {Cominsky}, {Conrad}, {Dermer}, {de Palma}, {Digel}, {Silva}, {Drell},
  {Dubois}, {Dumora}, {Favuzzi}, {Fegan}, {Ferrara}, {Frailis}, {Fukazawa},
  {Fusco}, {Gargano}, {Gehrels}, {Germani}, {Giglietto}, {Giordano}, {Godfrey},
  {Grenier}, {Grondin}, {Grove}, {Guillemot}, {Guiriec}, {Hadasch}, {Hanabata},
  {Harding}, {Hayashida}, {Hays}, {Hill}, {Horan}, {Hughes}, {Itoh}, {Jackson},
  {J{\'o}hannesson}, {Johnson}, {Johnson}, {Kamae}, {Katagiri}, {Kataoka},
  {Kerr}, {Kn{\"o}dlseder}, {Kuss}, {Lande}, {Latronico}, {Lee},
  {Lemoine-Goumard}, {Livingstone}, {Llena Garde}, {Longo}, {Loparco},
  {Lovellette}, {Lubrano}, {Makeev}, {Mazziotta}, {McEnery}, {Mehault},
  {Michelson}, {Mitthumsiri}, {Mizuno}, {Moiseev}, {Monte}, {Monzani},
  {Morselli}, {Moskalenko}, {Murgia}, {Nakamori}, {Naumann-Godo}, {Nolan},
  {Norris}, {Nuss}, {Ohsugi}, {Okumura}, {Omodei}, {Orlando}, {Ormes}, {Ozaki},
  {Panetta}, {Parent}, {Pelassa}, {Pepe}, {Pesce-Rollins}, {Piron}, {Porter},
  {Rain{\`o}}, {Rando}, {Razzano}, {Reimer}, {Reimer}, {Reposeur}, {Rodriguez},
  {Romani}, {Roth}, {Sadrozinski}, {Sander}, {Saz Parkinson}, {Scargle},
  {Sgr{\`o}}, {Siskind}, {Smith}, {Smith}, {Spandre}, {Spinelli}, {Strickman},
  {Suson}, {Takahashi}, {Takahashi}, {Tanaka}, {Thayer}, {Thayer}, {Thompson},
  {Tibaldo}, {Tibolla}, {Torres}, {Tosti}, {Tramacere}, {Uchiyama}, {Usher},
  {Vandenbroucke}, {Vasileiou}, {Vilchez}, {Vitale}, {Waite}, {Wallace},
  {Wang}, {Winer}, {Wood}, {Yang}, {Ylinen}, \& {Ziegler}}]{abdo2010}
{Abdo}, A.~A., {Ackermann}, M., {Ajello}, M., {et~al.} 2010, \apj, 723, 649

\bibitem[{{Abdo} {et~al.}(2011){Abdo}, {Ackermann}, {Ajello}, {Allafort},
  {Ballet}, {Barbiellini}, {Bastieri}, {Bechtol}, {Bellazzini}, {Berenji},
  {Blandford}, {Bonamente}, {Borgland}, {Bregeon}, {Brigida}, {Bruel},
  {Buehler}, {Buson}, {Caliandro}, {Cameron}, {Camilo}, {Caraveo}, {Cecchi},
  {Charles}, {Chaty}, {Chekhtman}, {Chernyakova}, {Cheung}, {Chiang},
  {Ciprini}, {Claus}, {Cohen-Tanugi}, {Cominsky}, {Corbel}, {Cutini},
  {D'Ammando}, {de Angelis}, {den Hartog}, {de Palma}, {Dermer}, {Digel},
  {Silva}, {Dormody}, {Drell}, {Drlica-Wagner}, {Dubois}, {Dubus}, {Dumora},
  {Enoto}, {Espinoza}, {Favuzzi}, {Fegan}, {Ferrara}, {Focke}, {Fortin},
  {Fukazawa}, {Funk}, {Fusco}, {Gargano}, {Gasparrini}, {Gehrels}, {Germani},
  {Giglietto}, {Giommi}, {Giordano}, {Giroletti}, {Glanzman}, {Godfrey},
  {Grenier}, {Grondin}, {Grove}, {Grundstrom}, {Guiriec}, {Gwon}, {Hadasch},
  {Harding}, {Hayashida}, {Hays}, {J{\'o}hannesson}, {Johnson}, {Johnson},
  {Johnston}, {Kamae}, {Katagiri}, {Kataoka}, {Keith}, {Kerr},
  {Kn{\"o}dlseder}, {Kramer}, {Kuss}, {Lande}, {Lee}, {Lemoine-Goumard},
  {Longo}, {Loparco}, {Lovellette}, {Lubrano}, {Manchester}, {Marelli},
  {Mazziotta}, {Michelson}, {Mitthumsiri}, {Mizuno}, {Moiseev}, {Monte},
  {Monzani}, {Morselli}, {Moskalenko}, {Murgia}, {Nakamori}, {Naumann-Godo},
  {Neronov}, {Nolan}, {Norris}, {Noutsos}, {Nuss}, {Ohsugi}, {Okumura},
  {Omodei}, {Orlando}, {Paneque}, {Parent}, {Pesce-Rollins}, {Pierbattista},
  {Piron}, {Porter}, {Possenti}, {Rain{\`o}}, {Rando}, {Ray}, {Razzano},
  {Razzaque}, {Reimer}, {Reimer}, {Reposeur}, {Ritz}, {Sadrozinski}, {Scargle},
  {Sgr{\`o}}, {Shannon}, {Siskind}, {Smith}, {Spandre}, {Spinelli},
  {Strickman}, {Suson}, {Takahashi}, {Tanaka}, {Thayer}, {Thayer}, {Thompson},
  {Thorsett}, {Tibaldo}, {Tibolla}, {Torres}, {Tosti}, {Troja}, {Uchiyama},
  {Usher}, {Vandenbroucke}, {Vasileiou}, {Vianello}, {Vitale}, {Waite}, {Wang},
  {Winer}, {Wolff}, {Wood}, {Wood}, {Yang}, {Ziegler}, \& {Zimmer}}]{abdo2011}
{Abdo}, A.~A., {Ackermann}, M., {Ajello}, M., {et~al.} 2011, \apjl, 736, L11

\bibitem[{{Aharonian} {et~al.}(2005{\natexlab{a}}){Aharonian}, {Akhperjanian},
  {Aye}, {Bazer-Bachi}, {Beilicke}, {Benbow}, {Berge}, {Berghaus},
  {Bernl{\"o}hr}, {Boisson}, {Bolz}, {Borrel}, {Braun}, {Breitling}, {Brown},
  {Gordo}, {Chadwick}, {Chounet}, {Cornils}, {Costamante}, {Degrange},
  {Dickinson}, {Djannati-Ata{\"i}}, {Drury}, {Dubus}, {Emmanoulopoulos},
  {Espigat}, {Feinstein}, {Fleury}, {Fontaine}, {Fuchs}, {Funk}, {Gallant},
  {Giebels}, {Gillessen}, {Glicenstein}, {Goret}, {Hadjichristidis}, {Hauser},
  {Heinzelmann}, {Henri}, {Hermann}, {Hinton}, {Hofmann}, {Holleran}, {Horns},
  {Jacholkowska}, {de Jager}, {Kh{\'e}lifi}, {Komin}, {Konopelko}, {Latham},
  {Le Gallou}, {Lemi{\`e}re}, {Lemoine-Goumard}, {Leroy}, {Lohse}, {Marcowith},
  {Martin}, {Martineau-Huynh}, {Masterson}, {McComb}, {de Naurois}, {Nolan},
  {Noutsos}, {Orford}, {Osborne}, {Ouchrif}, {Panter}, {Pelletier}, {Pita},
  {P{\"u}hlhofer}, {Punch}, {Raubenheimer}, {Raue}, {Raux}, {Rayner}, {Reimer},
  {Reimer}, {Ripken}, {Rob}, {Rolland}, {Rowell}, {Sahakian}, {Saug{\'e}},
  {Schlenker}, {Schlickeiser}, {Schuster}, {Schwanke}, {Siewert}, {Sol},
  {Spangler}, {Steenkamp}, {Stegmann}, {Tavernet}, {Terrier}, {Th{\'e}oret},
  {Tluczykont}, {Vasileiadis}, {Venter}, {Vincent}, {V{\"o}lk}, \&
  {Wagner}}]{aharonian2005ls5039}
{Aharonian}, F., {Akhperjanian}, A.~G., {Aye}, K.-M., {et~al.}
  2005{\natexlab{a}}, Science, 309, 746

\bibitem[{{Aharonian} {et~al.}(2005{\natexlab{b}}){Aharonian}, {Akhperjanian},
  {Aye}, {Bazer-Bachi}, {Beilicke}, {Benbow}, {Berge}, {Berghaus},
  {Bernl{\"o}hr}, {Boisson}, {Bolz}, {Braun}, {Breitling}, {Brown}, {Bussons
  Gordo}, {Chadwick}, {Chounet}, {Cornils}, {Costamante}, {Degrange},
  {Djannati-Ata{\"i}}, {O'C.~Drury}, {Dubus}, {Emmanoulopoulos}, {Espigat},
  {Feinstein}, {Fleury}, {Fontaine}, {Fuchs}, {Funk}, {Gallant}, {Giebels},
  {Gillessen}, {Glicenstein}, {Goret}, {Hadjichristidis}, {Hauser},
  {Heinzelmann}, {Henri}, {Hermann}, {Hinton}, {Hofmann}, {Holleran}, {Horns},
  {de Jager}, {Johnston}, {Kh{\'e}lifi}, {Kirk}, {Komin}, {Konopelko},
  {Latham}, {Le Gallou}, {Lemi{\`e}re}, {Lemoine-Goumard}, {Leroy},
  {Martineau-Huynh}, {Lohse}, {Marcowith}, {Masterson}, {McComb}, {de Naurois},
  {Nolan}, {Noutsos}, {Orford}, {Osborne}, {Ouchrif}, {Panter}, {Pelletier},
  {Pita}, {P{\"u}hlhofer}, {Punch}, {Raubenheimer}, {Raue}, {Raux}, {Rayner},
  {Redondo}, {Reimer}, {Reimer}, {Ripken}, {Rob}, {Rolland}, {Rowell},
  {Sahakian}, {Saug{\'e}}, {Schlenker}, {Schlickeiser}, {Schuster}, {Schwanke},
  {Siewert}, {Skj{\ae}raasen}, {Sol}, {Steenkamp}, {Stegmann}, {Tavernet},
  {Terrier}, {Th{\'e}oret}, {Tluczykont}, {Vasileiadis}, {Venter}, {Vincent},
  {V{\"o}lk}, \& {Wagner}}]{aharonian2005psr}
{Aharonian}, F., {Akhperjanian}, A.~G., {Aye}, K.-M., {et~al.}
  2005{\natexlab{b}}, \aap, 442, 1

\bibitem[{{Albert} {et~al.}(2006){Albert}, {Aliu}, {Anderhub}, {Antoranz},
  {Armada}, {Asensio}, {Baixeras}, {Barrio}, {Bartelt}, {Bartko}, {Bastieri},
  {Bavikadi}, {Bednarek}, {Berger}, {Bigongiari}, {Biland}, {Bisesi}, {Bock},
  {Bordas}, {Bosch-Ramon}, {Bretz}, {Britvitch}, {Camara}, {Carmona},
  {Chilingarian}, {Ciprini}, {Coarasa}, {Commichau}, {Contreras}, {Cortina},
  {Curtef}, {Danielyan}, {Dazzi}, {De Angelis}, {de los Reyes}, {De Lotto},
  {Domingo-Santamar{\'{\i}}a}, {Dorner}, {Doro}, {Errando}, {Fagiolini},
  {Ferenc}, {Fern{\'a}ndez}, {Firpo}, {Flix}, {Fonseca}, {Font}, {Fuchs},
  {Galante}, {Garczarczyk}, {Gaug}, {Giller}, {Goebel}, {Hakobyan},
  {Hayashida}, {Hengstebeck}, {H{\"o}hne}, {Hose}, {Hsu}, {Isar}, {Jacon},
  {Kalekin}, {Kosyra}, {Kranich}, {Laatiaoui}, {Laille}, {Lenisa}, {Liebing},
  {Lindfors}, {Lombardi}, {Longo}, {L{\'o}pez}, {L{\'o}pez}, {Lorenz},
  {Lucarelli}, {Majumdar}, {Maneva}, {Mannheim}, {Mansutti}, {Mariotti},
  {Mart{\'{\i}}nez}, {Mase}, {Mazin}, {Merck}, {Meucci}, {Meyer}, {Miranda},
  {Mirzoyan}, {Mizobuchi}, {Moralejo}, {Nilsson}, {O{\~n}a-Wilhelmi},
  {Ordu{\~n}a}, {Otte}, {Oya}, {Paneque}, {Paoletti}, {Paredes}, {Pasanen},
  {Pascoli}, {Pauss}, {Pavel}, {Pegna}, {Persic}, {Peruzzo}, {Piccioli},
  {Poller}, {Pooley}, {Prandini}, {Raymers}, {Rhode}, {Rib{\'o}}, {Rico},
  {Riegel}, {Rissi}, {Robert}, {Romero}, {R{\"u}gamer}, {Saggion},
  {S{\'a}nchez}, {Sartori}, {Scalzotto}, {Scapin}, {Schmitt}, {Schweizer},
  {Shayduk}, {Shinozaki}, {Shore}, {Sidro}, {Sillanp{\"a}{\"a}}, {Sobczynska},
  {Stamerra}, {Stark}, {Takalo}, {Temnikov}, {Tescaro}, {Teshima}, {Tonello},
  {Torres}, {Torres}, {Turini}, {Vankov}, {Vitale}, {Wagner}, {Wibig},
  {Wittek}, {Zanin}, \& {Zapatero}}]{albert2006}
{Albert}, J., {Aliu}, E., {Anderhub}, H., {et~al.} 2006, Science, 312, 1771

\bibitem[{{An} {et~al.}(2015){An}, {Bellm}, {Bhalerao}, {Boggs}, {Christensen},
  {Craig}, {Fuerst}, {Hailey}, {Harrison}, {Kaspi}, {Natalucci}, {Stern},
  {Tomsick}, \& {Zhang}}]{an2015}
{An}, H., {Bellm}, E., {Bhalerao}, V., {et~al.} 2015, \apj, 806, 166

\bibitem[{{An} \& {Romani}(2017)}]{an2017}
{An}, H. \& {Romani}, R.~W. 2017, \apj, 838, 145

\bibitem[{{Astropy Collaboration} {et~al.}(2013){Astropy Collaboration},
  {Robitaille}, {Tollerud}, {Greenfield}, {Droettboom}, {Bray}, {Aldcroft},
  {Davis}, {Ginsburg}, {Price-Whelan}, {Kerzendorf}, {Conley}, {Crighton},
  {Barbary}, {Muna}, {Ferguson}, {Grollier}, {Parikh}, {Nair}, {Unther},
  {Deil}, {Woillez}, {Conseil}, {Kramer}, {Turner}, {Singer}, {Fox}, {Weaver},
  {Zabalza}, {Edwards}, {Azalee Bostroem}, {Burke}, {Casey}, {Crawford},
  {Dencheva}, {Ely}, {Jenness}, {Labrie}, {Lim}, {Pierfederici}, {Pontzen},
  {Ptak}, {Refsdal}, {Servillat}, \& {Streicher}}]{astropy2013}
{Astropy Collaboration}, {Robitaille}, T.~P., {Tollerud}, E.~J., {et~al.} 2013,
  \aap, 558, A33

\bibitem[{{Bailer-Jones}(2015)}]{bailerjones2015}
{Bailer-Jones}, C.~A.~L. 2015, \pasp, 127, 994

\bibitem[{{Beasley} {et~al.}(2002){Beasley}, {Gordon}, {Peck}, {Petrov},
  {MacMillan}, {Fomalont}, \& {Ma}}]{beasley2002}
{Beasley}, A.~J., {Gordon}, D., {Peck}, A.~B., {et~al.} 2002, \apjs, 141, 13

\bibitem[{{Camilo} {et~al.}(2001){Camilo}, {Bell}, {Manchester}, {Lyne},
  {Possenti}, {Kramer}, {Kaspi}, {Stairs}, {D'Amico}, {Hobbs}, {Gotthelf}, \&
  {Gaensler}}]{camilo2001}
{Camilo}, F., {Bell}, J.~F., {Manchester}, R.~N., {et~al.} 2001, \apjl, 557,
  L51

\bibitem[{{Camilo} {et~al.}(2004){Camilo}, {Gaensler}, {Gotthelf}, {Halpern},
  \& {Manchester}}]{camilo2004}
{Camilo}, F., {Gaensler}, B.~M., {Gotthelf}, E.~V., {Halpern}, J.~P., \&
  {Manchester}, R.~N. 2004, \apj, 616, 1118

\bibitem[{{Chatterjee} {et~al.}(2005){Chatterjee}, {Vlemmings}, {Brisken},
  {Lazio}, {Cordes}, {Goss}, {Thorsett}, {Fomalont}, {Lyne}, \&
  {Kramer}}]{chatterjee2005}
{Chatterjee}, S., {Vlemmings}, W.~H.~T., {Brisken}, W.~F., {et~al.} 2005,
  \apjl, 630, L61

\bibitem[{{Chen} {et~al.}(2017){Chen}, {Ng}, {Takata}, {Yu}, \&
  {Cheng}}]{chen2017}
{Chen}, A.~M., {Ng}, C.~W., {Takata}, J., {Yu}, Y.~W., \& {Cheng}, K.~S. 2017,
  ArXiv e-prints [\eprint[arXiv]{1703.08080}]

\bibitem[{{Coley}(2015)}]{coley2015}
{Coley}, J.~B. 2015, PhD thesis, University of Maryland, Baltimore County

\bibitem[{{Corbet} {et~al.}(2011){Corbet}, {Cheung}, {Kerr}, {Dubois},
  {Donato}, {Caliandro}, {Coe}, {Edwards}, {Filipovic}, {Payne}, \&
  {Stevens}}]{corbet2011}
{Corbet}, R.~H.~D., {Cheung}, C.~C., {Kerr}, M., {et~al.} 2011, The
  Astronomer's Telegram, 3221, 1

\bibitem[{{Corbet} {et~al.}(2016){Corbet}, {Chomiuk}, {Coe}, {Coley}, {Dubus},
  {Edwards}, {Martin}, {McBride}, {Stevens}, {Strader}, {Townsend}, \&
  {Udalski}}]{corbet2016}
{Corbet}, R.~H.~D., {Chomiuk}, L., {Coe}, M.~J., {et~al.} 2016, \apj, 829, 105

\bibitem[{{Dubus}(2013)}]{dubus2013}
{Dubus}, G. 2013, \aapr, 21, 64

\bibitem[{{Fermi LAT Collaboration} {et~al.}(2012){Fermi LAT Collaboration},
  {Ackermann}, {Ajello}, {Ballet}, {Barbiellini}, {Bastieri}, {Belfiore},
  {Bellazzini}, {Berenji}, {Blandford}, {Bloom}, {Bonamente}, {Borgland},
  {Bregeon}, {Brigida}, {Bruel}, {Buehler}, {Buson}, {Caliandro}, {Cameron},
  {Caraveo}, {Cavazzuti}, {Cecchi}, {{\c C}elik}, {Charles}, {Chaty},
  {Chekhtman}, {Cheung}, {Chiang}, {Ciprini}, {}, {Claus}, {Cohen-Tanugi},
  {Corbel}, {Corbet}, {Cutini}, {de Luca}, {den Hartog}, {de Palma}, {Dermer},
  {Digel}, {do Couto e Silva}, {Donato}, {Drell}, {Drlica-Wagner}, {Dubois},
  {Dubus}, {Favuzzi}, {Fegan}, {Ferrara}, {Focke}, {Fortin}, {Fukazawa},
  {Funk}, {Fusco}, {Gargano}, {Gasparrini}, {Gehrels}, {Germani}, {Giglietto},
  {Giordano}, {Giroletti}, {Glanzman}, {Godfrey}, {Grenier}, {Grove},
  {Guiriec}, {Hadasch}, {Hanabata}, {Harding}, {Hayashida}, {Hays}, {Hill},
  {Hughes}, {J{\'o}hannesson}, {Johnson}, {Johnson}, {Kamae}, {Katagiri},
  {Kataoka}, {Kerr}, {Kn{\"o}dlseder}, {Kuss}, {Lande}, {Longo}, {Loparco},
  {Lovellette}, {Lubrano}, {Mazziotta}, {McEnery}, {Michelson}, {Mitthumsiri},
  {Mizuno}, {Monte}, {Monzani}, {Morselli}, {Moskalenko}, {Murgia}, {Nakamori},
  {Naumann-Godo}, {Norris}, {Nuss}, {Ohno}, {Ohsugi}, {Okumura}, {Omodei},
  {Orlando}, {Ozaki}, {Paneque}, {Parent}, {Pesce-Rollins}, {Pierbattista},
  {Piron}, {Pivato}, {Porter}, {Rain{\`o}}, {Rando}, {Razzano}, {Reimer},
  {Reimer}, {Ritz}, {Romani}, {Roth}, {Saz Parkinson}, {Sgr{\`o}}, {Siskind},
  {Spandre}, {Spinelli}, {Suson}, {Takahashi}, {Tanaka}, {Thayer}, {Thayer},
  {Thompson}, {Tibaldo}, {Tinivella}, {Torres}, {Tosti}, {Troja}, {Uchiyama},
  {Usher}, {Vandenbroucke}, {Vianello}, {Vitale}, {Waite}, {Winer}, {Wood},
  {Wood}, {Yang}, {Zimmer}, {Coe}, {Di Mille}, {Edwards}, {Filipovi{\'c}},
  {Payne}, {Stevens}, \& {Torres}}]{fermi2012}
{Fermi LAT Collaboration}, {Ackermann}, M., {Ajello}, M., {et~al.} 2012,
  Science, 335, 189

\bibitem[{{Gaia Collaboration} {et~al.}(2018){Gaia Collaboration}, {Brown},
  {Vallenari}, {Prusti}, {de Bruijne}, {Babusiaux}, \&
  {Bailer-Jones}}]{gaia2018}
{Gaia Collaboration}, {Brown}, A.~G.~A., {Vallenari}, A., {et~al.} 2018, ArXiv
  e-prints [\eprint[arXiv]{1804.09365}]

\bibitem[{{Gaia Collaboration} {et~al.}(2016){Gaia Collaboration}, {Brown},
  {Vallenari}, {Prusti}, {de Bruijne}, {Mignard}, {Drimmel}, {Babusiaux},
  {Bailer-Jones}, {Bastian}, \& et~al.}]{gaia2016}
{Gaia Collaboration}, {Brown}, A.~G.~A., {Vallenari}, A., {et~al.} 2016, \aap,
  595, A2

\bibitem[{{G{\'o}mez} {et~al.}(2008){G{\'o}mez}, {Rodr{\'{\i}}guez}, {Loinard},
  {Lizano}, {Allen}, {Poveda}, \& {Menten}}]{gomez2008}
{G{\'o}mez}, L., {Rodr{\'{\i}}guez}, L.~F., {Loinard}, L., {et~al.} 2008, \apj,
  685, 333

\bibitem[{{Gordon} {et~al.}(2016){Gordon}, {Jacobs}, {Beasley}, {Peck},
  {Gaume}, {Charlot}, {Fey}, {Ma}, {Titov}, \& {Boboltz}}]{gordon2016}
{Gordon}, D., {Jacobs}, C., {Beasley}, A., {et~al.} 2016, \aj, 151, 154

\bibitem[{{Greisen}(2003)}]{greisen2003}
{Greisen}, E.~W. 2003, in Astrophysics and Space Science Library, Vol. 285,
  Information Handling in Astronomy - Historical Vistas, ed. A.~{Heck}, 109

\bibitem[{{H.~E.~S.~S.~Collaboration}
  {et~al.}(2012){H.~E.~S.~S.~Collaboration}, {Abramowski}, {Acero},
  {Aharonian}, {Akhperjanian}, {Anton}, {Balzer}, {Barnacka}, {Becherini},
  {Becker}, {Bernl{\"o}h}, {Birsin}, {Biteau}, {Bochow}, {Boisson}, {Bolmont},
  {Bordas}, {Brucker}, {Brun}, {Brun}, {Bulik}, {B{\"u}sching}, {Carrigan},
  {Casanova}, {Cerruti}, {Chadwick}, {Charbonnier}, {Chaves}, {Cheesebrough},
  {Cologna}, {Conrad}, {Dalton}, {Daniel}, {Davids}, {Degrange}, {Deil},
  {Dickinson}, {Djannati-Ata{\"i}}, {Domainko}, {Drury}, {Dubus}, {Dutson},
  {Dyks}, {Dyrda}, {Egberts}, {Eger}, {Espigat}, {Fallon}, {Fegan},
  {Feinstein}, {Fernandes}, {Fiasson}, {Fontaine}, {F{\"o}rster},
  {F{\"u}{\ss}ling}, {Gallant}, {Gast}, {G{\'e}rard}, {Gerbig}, {Giebels},
  {Glicenstein}, {Gl{\"u}ck}, {G{\"o}ring}, {H{\"a}ffner}, {Hague}, {Hahn},
  {Hampf}, {Harris}, {Hauser}, {Heinz}, {Heinzelmann}, {Henri}, {Hermann},
  {Hillert}, {Hinton}, {Hofmann}, {Hofverberg}, {Holler}, {Horns},
  {Jacholkowska}, {de Jager}, {Jahn}, {Jamrozy}, {Jung}, {Kastendieck},
  {Katarzy{\'n}ski}, {Katz}, {Kaufmann}, {Keogh}, {Kh{\'e}lifi}, {Klochkov},
  {Klu{\.z}niak}, {Kneiske}, {Komin}, {Kosack}, {Kossakowski}, {Krayzel},
  {Laffon}, {Lamanna}, {Lenain}, {Lennarz}, {Lohse}, {Lopatin}, {Lu},
  {Marandon}, {Marcowith}, {Masbou}, {Maxted}, {Mayer}, {McComb}, {Medina},
  {M{\'e}hault}, {Moderski}, {Mohamed}, {Moulin}, {Naumann}, {Naumann-Godo},
  {de Naurois}, {Nedbal}, {Nekrassov}, {Nguyen}, {Nicholas}, {Niemiec},
  {Nolan}, {Ohm}, {de O{\~n}a Wilhelmi}, {Opitz}, {Ostrowski}, {Oya}, {Panter},
  {Paz Arribas}, {Pekeur}, {Pelletier}, {Perez}, {Petrucci}, {Peyaud}, {Pita},
  {P{\"u}hlhofer}, {Punch}, {Quirrenbach}, {Raue}, {Rayner}, {Reimer},
  {Reimer}, {Renaud}, {de los Reyes}, {Rieger}, {Ripken}, {Rob}, {Rosier-Lees},
  {Rowell}, {Rudak}, {Rulten}, {Sahakian}, {Sanchez}, {Santangelo},
  {Schlickeiser}, {Schulz}, {Schwanke}, {Schwarzburg}, {Schwemmer}, {Sheidaei},
  {Skilton}, {Sol}, {Spengler}, {Stawarz}, {Steenkamp}, {Stegmann}, {Stinzing},
  {Stycz}, {Sushch}, {Szostek}, {Tavernet}, {Terrier}, {Tluczykont},
  {Valerius}, {van Eldik}, {Vasileiadis}, {Venter}, {Viana}, {Vincent},
  {V{\"o}lk}, {Volpe}, {Vorobiov}, {Vorster}, {Wagner}, {Ward}, {White},
  {Wierzcholska}, {Zacharias}, {Zajczyk}, {Zdziarski}, {Zech}, \&
  {Zechlin}}]{hess2012}
{H.~E.~S.~S.~Collaboration}, {Abramowski}, A., {Acero}, F., {et~al.} 2012,
  \aap, 541, A5

\bibitem[{{H.~E.~S.~S.~Collaboration}
  {et~al.}(2015){H.~E.~S.~S.~Collaboration}, {Abramowski}, {Aharonian}, {Ait
  Benkhali}, {Akhperjanian}, {Ang{\"u}ner}, {Backes}, {Balzer}, {Becherini},
  {Becker Tjus}, \& et~al.}]{hess2015}
{H.~E.~S.~S.~Collaboration}, {Abramowski}, A., {Aharonian}, F., {et~al.} 2015,
  \aap, 577, A131

\bibitem[{{Hinton} {et~al.}(2009){Hinton}, {Skilton}, {Funk}, {Brucker},
  {Aharonian}, {Dubus}, {Fiasson}, {Gallant}, {Hofmann}, {Marcowith}, \&
  {Reimer}}]{hinton2009}
{Hinton}, J.~A., {Skilton}, J.~L., {Funk}, S., {et~al.} 2009, \apjl, 690, L101

\bibitem[{Hunter(2007)}]{hunter2007}
Hunter, J.~D. 2007, Computing In Science \& Engineering, 9, 90

\bibitem[{{Leser} {et~al.}(2017){Leser}, {Ohm}, {F{\"u}{\ss}ling}, {de
  Naurois}, {Egberts}, {Bordas}, {Klepser}, {Reimer}, {Reimer}, {Hinton}, \&
  {for the H.~E.~S.~S.~collaboration}}]{leser2017}
{Leser}, E., {Ohm}, S., {F{\"u}{\ss}ling}, M., {et~al.} 2017, ArXiv e-prints
  [\eprint[arXiv]{1708.01033}]

\bibitem[{{Luri} {et~al.}(2018){Luri}, {Brown}, {Sarro}, {Arenou},
  {Bailer-Jones}, {Castro-Ginard}, {de Bruijne}, {Prusti}, {Babusiaux}, \&
  {Delgado}}]{luri2018}
{Luri}, X., {Brown}, A.~G.~A., {Sarro}, L.~M., {et~al.} 2018, ArXiv e-prints
  [\eprint[arXiv]{1804.09376}]

\bibitem[{{Lyne} {et~al.}(2015){Lyne}, {Stappers}, {Keith}, {Ray}, {Kerr},
  {Camilo}, \& {Johnson}}]{lyne2015}
{Lyne}, A.~G., {Stappers}, B.~W., {Keith}, M.~J., {et~al.} 2015, \mnras, 451,
  581

\bibitem[{{Marcote}(2016)}]{marcote2015thesis}
{Marcote}, B. 2016, PhD thesis, University of Barcelona

\bibitem[{{Marcote} {et~al.}(2015){Marcote}, {Rib{\'o}}, {Paredes}, \&
  {Ishwara-Chandra}}]{marcote2015ls5039}
{Marcote}, B., {Rib{\'o}}, M., {Paredes}, J.~M., \& {Ishwara-Chandra}, C.~H.
  2015, \mnras, 451, 4578

\bibitem[{{Marcote} {et~al.}(2016){Marcote}, {Rib{\'o}}, {Paredes},
  {Ishwara-Chandra}, {Swinbank}, {Broderick}, {Markoff}, {Fender}, {Wijers},
  {Pooley}, {Stewart}, {Bell}, {Breton}, {Carbone}, {Corbel}, {Eisl\"offel},
  {Falcke}, {Grie{\ss}meier}, {Kuniyoshi}, {Pietka}, {Rowlinson}, {Serylak},
  {van der Horst}, {van Leeuwen}, {Wise}, \& {Zarka}}]{marcote2016}
{Marcote}, B., {Rib{\'o}}, M., {Paredes}, J.~M., {et~al.} 2016

\bibitem[{{Mart{\'{\i}}-Vidal} {et~al.}(2012){Mart{\'{\i}}-Vidal},
  {P{\'e}rez-Torres}, \& {Lobanov}}]{martividal2012}
{Mart{\'{\i}}-Vidal}, I., {P{\'e}rez-Torres}, M.~A., \& {Lobanov}, A.~P. 2012,
  \aap, 541, A135

\bibitem[{{Massey} \& {Meyer}(2001)}]{massey2001}
{Massey}, P. \& {Meyer}, M. 2001, {Stellar Masses}, ed. P.~{Murdin}

\bibitem[{{Mauch} {et~al.}(2003){Mauch}, {Murphy}, {Buttery}, {Curran},
  {Hunstead}, {Piestrzynski}, {Robertson}, \& {Sadler}}]{mauch2003}
{Mauch}, T., {Murphy}, T., {Buttery}, H.~J., {et~al.} 2003, \mnras, 342, 1117

\bibitem[{{Mel'nik} \& {Dambis}(2017)}]{melnik2017}
{Mel'nik}, A.~M. \& {Dambis}, A.~K. 2017, \mnras, 472, 3887

\bibitem[{{Miller-Jones} {et~al.}(2018){Miller-Jones}, {Deller}, {Shannon},
  {Dodson}, {Mold{\'o}n}, {Rib{\'o}}, {Dubus}, {Johnston}, {Paredes}, {Ransom},
  \& {Tomsick}}]{millerjones2018}
{Miller-Jones}, J.~C.~A., {Deller}, A.~T., {Shannon}, R.~M., {et~al.} 2018,
  \mnras, 479, 4849

\bibitem[{{Mills} {et~al.}(1961){Mills}, {Slee}, \& {Hill}}]{mills1961}
{Mills}, B.~Y., {Slee}, O.~B., \& {Hill}, E.~R. 1961, Australian Journal of
  Physics, 14, 497

\bibitem[{{Milne} {et~al.}(1989){Milne}, {Caswell}, {Kesteven}, {Haynes}, \&
  {Roger}}]{milne1989}
{Milne}, D.~K., {Caswell}, J.~L., {Kesteven}, M.~J., {Haynes}, R.~F., \&
  {Roger}, R.~S. 1989, Proceedings of the Astronomical Society of Australia, 8,
  187

\bibitem[{{Milne} \& {Dickel}(1975)}]{milne1975}
{Milne}, D.~K. \& {Dickel}, J.~R. 1975, Australian Journal of Physics, 28, 209

\bibitem[{{Mold{\'o}n}(2012)}]{moldon2012thesis}
{Mold{\'o}n}, J. 2012, PhD thesis, Universitat de Barcelona

\bibitem[{{Mold{\'o}n} {et~al.}(2012{\natexlab{a}}){Mold{\'o}n}, {Rib\'o}, \&
  {Paredes}}]{moldon2012ls5039}
{Mold{\'o}n}, J., {Rib\'o}, M., \& {Paredes}, J.~M. 2012{\natexlab{a}}, \aap,
  548, A103

\bibitem[{{Mold{\'o}n} {et~al.}(2012{\natexlab{b}}){Mold{\'o}n}, {Rib{\'o}},
  {Paredes}, {Brisken}, {Dhawan}, {Kramer}, {Lyne}, \&
  {Stappers}}]{moldon2012lspsr}
{Mold{\'o}n}, J., {Rib{\'o}}, M., {Paredes}, J.~M., {et~al.}
  2012{\natexlab{b}}, \aap, 543, A26

\bibitem[{{Monageng} {et~al.}(2017){Monageng}, {McBride}, {Townsend},
  {Kniazev}, {Mohamed}, \& {B\"ottcher}}]{monageng2017}
{Monageng}, I.~M., {McBride}, V.~A., {Townsend}, L.~J., {et~al.} 2017, \apj,
  847, 68

\bibitem[{{Napoli} {et~al.}(2011){Napoli}, {McSwain}, {Boyer}, \&
  {Roettenbacher}}]{napoli2011}
{Napoli}, V.~J., {McSwain}, M.~V., {Boyer}, A.~N.~M., \& {Roettenbacher}, R.~M.
  2011, \pasp, 123, 1262

\bibitem[{{Nelemans} {et~al.}(1999){Nelemans}, {Tauris}, \& {van den
  Heuvel}}]{nelemans1999}
{Nelemans}, G., {Tauris}, T.~M., \& {van den Heuvel}, E.~P.~J. 1999, \aap, 352,
  L87

\bibitem[{{Pradel} {et~al.}(2006){Pradel}, {Charlot}, \&
  {Lestrade}}]{pradel2006}
{Pradel}, N., {Charlot}, P., \& {Lestrade}, J.-F. 2006, \aap, 452, 1099

\bibitem[{{Reid} {et~al.}(2014){Reid}, {Menten}, {Brunthaler}, {Zheng}, {Dame},
  {Xu}, {Wu}, {Zhang}, {Sanna}, {Sato}, {Hachisuka}, {Choi}, {Immer},
  {Moscadelli}, {Rygl}, \& {Bartkiewicz}}]{reid2014}
{Reid}, M.~J., {Menten}, K.~M., {Brunthaler}, A., {et~al.} 2014, \apj, 783, 130

\bibitem[{{Ruiz} \& {May}(1986)}]{ruiz1986}
{Ruiz}, M.~T. \& {May}, J. 1986, \apj, 309, 667

\bibitem[{{Sault} {et~al.}(1995){Sault}, {Teuben}, \& {Wright}}]{sault1995}
{Sault}, R.~J., {Teuben}, P.~J., \& {Wright}, M.~C.~H. 1995, in Astronomical
  Society of the Pacific Conference Series, Vol.~77, Astronomical Data Analysis
  Software and Systems IV, ed. R.~A. {Shaw}, H.~E. {Payne}, \& J.~J.~E.
  {Hayes}, 433

\bibitem[{{Shepherd} {et~al.}(1994){Shepherd}, {Pearson}, \&
  {Taylor}}]{shepherd1994}
{Shepherd}, M.~C., {Pearson}, T.~J., \& {Taylor}, G.~B. 1994, in Bulletin of
  the American Astronomical Society, Vol.~26, Bulletin of the American
  Astronomical Society, 987

\bibitem[{{Skilton} {et~al.}(2009){Skilton}, {Pandey-Pommier}, {Hinton},
  {Cheung}, {Aharonian}, {Brucker}, {Dubus}, {Fiasson}, {Funk}, {Gallant},
  {Marcowith}, \& {Reimer}}]{skilton2009}
{Skilton}, J.~L., {Pandey-Pommier}, M., {Hinton}, J.~A., {et~al.} 2009, \mnras,
  399, 317

\bibitem[{{Strader} {et~al.}(2015){Strader}, {Chomiuk}, {Cheung}, {Salinas}, \&
  {Peacock}}]{strader2015}
{Strader}, J., {Chomiuk}, L., {Cheung}, C.~C., {Salinas}, R., \& {Peacock}, M.
  2015, \apjl, 813, L26

\bibitem[{{Veritas} \& {MAGIC Collaborations}(2017)}]{veritasmagic2017}
{Veritas} \& {MAGIC Collaborations}. 2017, The Astronomer's Telegram, 10810, 1

\bibitem[{{Waisberg} \& {Romani}(2015)}]{waisberg2015}
{Waisberg}, I.~R. \& {Romani}, R.~W. 2015, \apj, 805, 18

\bibitem[{{Wakely} \& {Horan}(2008)}]{wakely2008}
{Wakely}, S.~P. \& {Horan}, D. 2008, International Cosmic Ray Conference, 3,
  1341

\bibitem[{{Williams} {et~al.}(2015){Williams}, {Rangelov}, {Kargaltsev}, \&
  {Pavlov}}]{williams2015}
{Williams}, B.~J., {Rangelov}, B., {Kargaltsev}, O., \& {Pavlov}, G.~G. 2015,
  \apjl, 808, L19

\bibitem[{{Wu} {et~al.}(2018){Wu}, {Torricelli-Ciamponi}, {Massi}, {Reid},
  {Zhang}, {Shao}, \& {Zheng}}]{wu2018}
{Wu}, Y.~W., {Torricelli-Ciamponi}, G., {Massi}, M., {et~al.} 2018, \mnras,
  474, 4245

\bibitem[{{Zacharias} {et~al.}(2013){Zacharias}, {Finch}, {Girard}, {Henden},
  {Bartlett}, {Monet}, \& {Zacharias}}]{zacharias2013}
{Zacharias}, N., {Finch}, C.~T., {Girard}, T.~M., {et~al.} 2013, \aj, 145, 44

\end{thebibliography}

\end{document}